\documentclass[10pt]{article}
\usepackage[tbtags]{amsmath}
\usepackage{hyperref}
\usepackage{amssymb,amsthm}
\usepackage{amsfonts}
\usepackage{graphicx}
\graphicspath{{figs/}}
\usepackage[left=2.5cm,right=2.2cm,top=0.5cm,bottom=1.3cm,includeheadfoot]{geometry}
\usepackage{indentfirst}
\usepackage{color}
\usepackage{cite}
\usepackage{mathtools}
\usepackage{graphicx}
\usepackage{subcaption}
\usepackage{tikz-cd}
\usepackage{enumitem}

\definecolor{cites}{rgb}{0.65, 0.20, 0.46}
\definecolor{chapters}{rgb}{0.25, 0.30, 0.56}
\definecolor{header}{rgb}{0.0, 0.5, 0.64}

\hypersetup{
	colorlinks=true, 
	linkcolor=header, 
	citecolor=cites,
	urlcolor=chapters
}

\usepackage[normalem]{ulem}
\DeclareMathOperator{\sech}{sech}

\newcommand{\rzero}{r_{\scalebox{0.65}{0}}}
\numberwithin{equation}{section}

\renewcommand*{\thefootnote}{\fnsymbol{footnote}}
\begin{document}
	\begin{center}
		{\Large\bf
			Asymptotics of Bianchi IX under the presence of matter: approximate Kasner map
		}
		\vskip 5mm
		{\large
			David Brizuela\footnote{Contact author: {\tt david.brizuela@ehu.eus}}
			and Sara F. Uria\footnote{Contact author: {\tt sara.fernandezu@ehu.eus}}
		}
		\vskip 3mm
		{\sl Department of Physics and EHU Quantum Center, University of the Basque Country UPV/EHU,\\
			Barrio Sarriena s/n, 48940 Leioa, Spain}\\
	\end{center}
	
	\setcounter{footnote}{0}
	\renewcommand*{\thefootnote}{\arabic{footnote}}
	
	\begin{abstract}
		The goal of this paper is to analyze the effects of the matter fields in the evolution of the Bianchi IX cosmology close to the singularity. Although the dynamics of this model is very involved, asymptotically, as the singularity is approached, it can be well approximated as a succession of Bianchi I periods connected by quick bounces against potential walls. Moreover, in such limit, matter fields (excluding stiff matter) are known to be subdominant with respect to the anisotropies. Therefore,
		by performing an expansion around small volumes, and assuming that the
		matter contribution in this regime can be described as a barotropic perfect fluid with a linear equation of state, we obtain an approximate analytic solution of the dynamics.
		Then, we explicitly compute the form of the transition (Kasner) map that relates the pre- and post-bounce Bianchi I periods including the leading matter effects. As an important conclusion, we observe that generically the presence of matter leads to a post-bounce velocity with a lower deflection angle than in the vacuum case, and thus effectively increases the convex curvature of the potential walls. This effect may have important consequences in the chaotic nature of general relativity near spacelike singularities.
	\end{abstract}

	\section{Introduction}
	\label{sec:intro}

	Singularities are a generic feature of the solutions of the Einstein equations.
	Even if general relativity ceases to be valid there, the study of the dynamics predicted by the theory near
	such singularities can be of key relevance to understand the behavior of the initial stages of the Universe or
	the interior of black holes. In this context, Belinski, Khalatnikov, and
	Lifshitz (BKL) \cite{Belinsky:1970ew} conjectured that,
	near spacelike singularities, spatial derivatives of the different physical quantities
	are negligible with respect to their time derivatives. That is, the dynamics is asymptotically local, and thus the evolution can be effectively
	described in terms of a homogeneous metric. Although there is not yet
	a strict mathematical proof of such, so-called BKL conjecture, there is ample of numerical evidence
	in its favor
	\cite{Berger:2002st,Berger:1993ff,Berger:1998vxa,Berger:1998wr,Berger:2014tev,Garfinkle:2003bb,Garfinkle:2004ww}.
	However, one should also mention that the presence of spikes (small-scale spatial structures)
	in certain numerical simulations casts doubts into the local nature of the conjecture \cite{Heinzle:2012um}.
	
	In this respect, the Bianchi models, which describe homogeneous, though generically anisotropic,
	spacetimes, stand out as paradigmatic examples to learn about general relativity in its most extreme regimes. In particular, the Bianchi IX spacetime has received special attention due
	to its rich dynamical properties and its generality: as a particular case, it contains the closed Friedmann-Lemaitre-Robertson-Walker (FLRW) cosmology,
	and, in different limits, the Bianchi I, II, as well as
	the flat FLRW models. In Refs. \cite{Misner:1969hg, Misner:1969ae},
	Misner analyzed this model in vacuum, and reached similar conclusions as BKL:
	near the singularity the dynamics of the universe can be mapped to the evolution
	of a particle moving on a potential that alternates periods of free evolution interrupted by
	bounces against certain potential walls. The map that relates the properties before and after the bounce
	is called the Kasner map (sometimes also the BKL map), and has been of paramount relevance in the analysis
	of the features of the model. Applying it recursively, it can be used to describe the evolution
	of the system toward the singularity in a discrete manner, and such analysis led to the conclusion
	that the model is chaotic \cite{Barrow:1981prl,Barrow:1981sx,Chernoff:1983zz}. The analysis of the chaotic nature
	of the full dynamics turned out to be very controversial \cite{Hobill:1993osv},
	especially due to the noninvariance of Lyapunov exponents
	under coordinate transformations. However, in Ref.~\cite{Motter:2003jm}
	it was clarified that, if the coordinates obey certain specific properties, the sign of the Lyapunov exponents is invariant.
	In this way, it was concluded that the full dynamics is indeed chaotic \cite{Imponente:2001fy}.
	An independent verification of such statement came from the study
	\cite{Cornish:1996yg, Cornish:1996hx} based on invariant fractal methods.
	Recently, considering a semiclassical regime, leading quantum-geometry
	effects were considered, and the corresponding generalized Kasner map was derived in Ref. \cite{Brizuela:2022uun}.
	In addition, in Refs. \cite{Bojowald:2023fas, Bojowald:2023sjw} the chaos of the quantum system was analyzed
	making use of both, the fractal method and the computation of the Lyapunov exponents in an adequate coordinate
	basis, with the conclusion that, even if the chaos
	persists, it is mitigated by quantum fluctuations.
	
	Concerning the coupling of matter to the Bianchi IX model, soon after the pioneering work
	by Misner~\cite{Misner:1969hg, Misner:1969ae}, in Ref.~\cite{Matzner:1970}, a dust field (pressureless fluid) was considered,
	and the evolution of anisotropies was extensively studied. In a series of papers
	\cite{Ryan:1971,Ryan:1971bis,Ryan:1972} (see also \cite{Ryanlectures, Ryan-Shepley}), making use of Hamiltonian techniques, Ryan
	systematically analyzed
	the most general (diagonal as well as nondiagonal) Bianchi IX spacetimes with a
	barotropic fluid, which could exhibit rotation, expansion, and shear.
	The main conclusion was that matter did not affect the qualitative behavior of
	the dynamics near the singularity, a feature that was also claimed
	by BKL \cite{Belinsky:1970ew}. This led to the statement that generically ``matter does not matter'',
	with the relevant exception of stiff matter (or, equivalently, a massless scalar field).
	The most precise formulation of this idea was presented in \cite{Ringstrom:2000mk} (see also Ref.~\cite{Heinzle:2009eh})
	in the form of a theorem,
	which states that, assuming an orthogonal perfect fluid with a linear
	equation of state, for stiff matter the solutions asymptotically
	converge to a point, while,
	for other matter types, the solutions converge to an attractor given by
	Bianchi II vacuum orbits.
	The body of results in this area is very extensive and we refer the reader to the reviews \cite{Wainwright, Jantzen:2001me},
	which include results for different Bianchi types,
	and to the critical report \cite{Heinzle:2009du} for a more detailed summary.
	In any case, even if matter fields are subdominant as compared to anisotropies
	as the system approaches the singularity, their effects are vanishing only
	in the exact limit of null volume.

	Therefore, the main goal
	of the present paper is to obtain the leading matter effects near the
	Bianchi IX singularity.
	For this purpose, we will assume that the matter content 
	in this region can be described as a barotropic perfect fluid with a linear equation of state.
	However, the equations are very involved and it turns out very difficult to get an analytic solution.
	Hence, making use of the fact that the equations of motion tend to the
	vacuum case as the system approaches the singularity,
	we will perform an expansion around the vacuum solution.
	Solving these approximate equations, we will
	be able to obtain the Kasner map that includes the leading matter effects.

The remaining of this paper is organized as follows. In Sec.~\ref{sec.bianchimodels} we present the diagonal Bianchi IX cosmology coupled to matter. Then, in Sec.~\ref{sec.BIX}, we explain the main assumptions that we will consider to solve the dynamics near the singularity, and also introduce
	the time gauge that will be used all along the paper. Once that this is set, in Sec.~\ref{sec.vacuum} the Kasner map for vacuum and stiff matter is obtained. This is review material, though we will present it in detail since it will serve as the basis for the subsequent study with a general barotropic fluid. Then, in Sec.~\ref{sec.generalkasner}, we present the main original results of this paper: by linearizing the equations around the previous vacuum solution, we analytically solve the Bianchi IX dynamics and obtain its corresponding Kasner map including the
	leading matter effects. Moreover, in Sec.~\ref{sec:curvatureBIX}, we analyze the asymptotics of the Kretschmann scalar. Finally, in Sec.~\ref{sec:conclusions}
	we summarize and discuss the main results of the paper.

	\section{The diagonal Bianchi IX model coupled to matter}\label{sec.bianchimodels}
	
	In the Bianchi classification of three-dimensional Lie algebras, the type IX corresponds to $\mathfrak{so}(3, \mathbb{R})$.
	The four-dimensional Bianchi IX geometry can then be constructed by considering spatial
	homogeneous sections with such isometry group, and assuming that
	the time vector generating the foliation into homogeneous slices is
	invariant under the action of the Killing fields.
	The resulting spacetime is of the form $\mathbb{R}\times S^3$,
	and, in adapted coordinates, the metric can be explicitly written as
	\begin{align}\label{metric}
	ds^2=-N^2(t) dt^2+\frac{\rzero^2}{4}\gamma_{ij}(t)\sigma^i\sigma^j,
	\end{align}
	where $N(t)$ is the lapse function, $\rzero$ is a constant with dimensions of length,
	and Latin indices go from 1 to 3. The $\sigma^i$ are the invariant one-forms that contain the information
	of the corresponding algebraic structure, and can be chosen as
	\begin{align}\label{dual_forms_1}
	\begin{aligned}	&\sigma^1:=\sin\psi
	d\theta-\cos\psi\sin\theta d\phi,
	\\
	&\sigma^2:=\cos\psi d\theta+\sin\psi\sin\theta d\phi,
	\\
	&\sigma^3:=-(d\psi+\cos\theta d\phi),
	\end{aligned}
	\end{align}
	with the angles $\theta$, $\phi$, and $\psi$.
	
	The time-dependent matrix $\gamma_{ij}$ is completely arbitrary but,
	following Misner \cite{Misner:1969hg, Misner:1969ae}, we will choose it to be diagonal and parametrized as 
	\begin{equation}
	\gamma_{ij}:={\rm diag}\left(a_1^2,a_2^2,a_3^2\right)={\rm diag}\left(e^{2(\alpha+\sqrt{3}\beta_-+\beta_+)},e^{2(\alpha-\sqrt{3}\beta_-+\beta_+)},e^{2(\alpha-2\beta_+)}\right).
	\end{equation}
	In particular, as will be commented below, this diagonal model does not allow for a single tilted fluid.
	For a recent discussion about the different dynamical behavior between this diagonal and the more general nondiagonal Bianchi IX model, we refer the reader to Refs. \cite{Czuchry:2014hxa,Kiefer:2018uyv,Kwidzinski:2019rwj}.
	The geometric interpretation of the variables $(\alpha, \beta_+,\beta_-)$
	can be readily seen by writing them in terms of the scale factors $(a_1, a_2, a_3)$:
	\begin{align}
	\label{def_misner_var}
	\alpha&=\frac{1}{3}\ln(a_1 a_2 a_3),\\
	\beta_{+}&=-\frac{1}{2}\ln \left[{\frac{a_3}{(a_1a_2a_3)^{1/3}}}\right],\\
	\beta_{-}&=\frac{1}{2\sqrt{3}}\ln \left({\frac{a_1}{a_2}}\right).
	\end{align}
	That is, the exponential $e^{3\alpha}$ encodes the volume of the spatial sections, while
	the shape parameters $\beta_+$ and $\beta_-$ provide a measure of the spatial anisotropy in the different directions.
	
	The evolution of the variables can then be obtained
	from the different components of the Einstein equations,
	\begin{equation}\label{Einstein}
	G^{\mu}{}_\nu=\kappa\, T^{\mu}{}_\nu,
	\end{equation}
	where the coupling constant $\kappa$ is given as
	$\kappa=8\pi G_N$, with $G_N$ being the Newton gravitational constant, and Greek indices run from 0 to 3.
	First, we note that, in the basis of the invariant one-forms $\{dt,\sigma^1,\sigma^2,\sigma^3\}$,
	the Einstein tensor $G^\mu{}_\nu$ corresponding to the metric \eqref{metric} is diagonal,
	and, thus, so must be the energy-momentum tensor,
	\begin{equation}\label{matterflux}
	 T^\mu{}_\nu=0,\quad {\rm for}\quad \mu\neq\nu.
	\end{equation}

Next, denoting with a prime the derivative with respect to the generic time $t$, from the ${00}$ component of the Einstein equations, one obtains the first-order (constraint) equation
\begin{equation}
	\label{eq_constraint}
	\kappa T^0{}_0+\frac{3}{N^2}\left({\alpha^{\prime}}^2-{\beta^\prime_+}^{\!\!2}-{\beta^{\prime}_-}^{\!\!2}\right)-\frac{6}{ \rzero^2}e^{-2 \alpha}U(\beta_+,\beta_-)=0,
	\end{equation}
	where we have defined the potential
	\begin{equation}
	\label{potential_Bianchi_IX}
	U(\beta_+,\beta_-):=\frac{1}{6}\left[
	e^{-8\beta_+}+2e^{4\beta_+}\left(
	\cosh(4\sqrt{3}\beta_-)-1
	\right)-4e^{-2\beta_+}\cosh(2\sqrt{3}\beta_-)
	\right],
	\end{equation}
	which is related to the three-dimensional Ricci scalar as ${}^{\tiny (3)}R=-12 U e^{-2\alpha}/\rzero^2$.
Finally, rearranging and simplifying the diagonal spatial components,
leads to the three second-order (evolution) equations,
\begin{align}
	\label{eq_motion_alpha}
		\alpha^{\prime\prime}&=-3{\alpha^{\prime}}^{2}+\frac{{N}^\prime{\alpha}^\prime}{N}+4N^2 \frac{e^{-2 \alpha}}{\rzero^2}\,U(\beta_+,\beta_-)
		-\frac{\kappa}{6}N^2(T+2T^0{}_0),\\
	\label{eq_motion_beta_+}
	{\beta}_+^{\prime\prime}&= -3\alpha^{\prime}\beta_+^{\prime} +\frac{N^{\prime}{\beta}_+^\prime}{N}-N^2\frac{e^{-2 \alpha}}{\rzero^2}\,\frac{\partial U(\beta_+,\beta_-)}{\partial\beta_+}+\frac{\kappa N^2}{6}(T^1{}_1+T^2{}_2-2 T^3{}_3),\\
	\label{eq_motion_beta_-}
	\beta_-^{\prime\prime}&= -3\alpha^{\prime}\beta^{\prime}_- +\frac{N^{\prime}\beta_-^{\prime}}{N}-N^2\frac{e^{-2 \alpha}}{\rzero^2}\,\frac{\partial U(\beta_+,\beta_-)}{\partial\beta_-}+\frac{\kappa N^2}{2\sqrt{3}}(T^1{}_1-T^2{}_2),
	\end{align}
	where $T:=T^\mu{}_\mu$ is the trace of the energy-momentum tensor.
It is interesting to note that,
a particular case of the above equations corresponds to the closed FLRW model.
	In such isotropic case the shape parameters vanish, $\beta_+=0$ and $\beta_-=0$, which leads to a constant value of the potential $U=-1/2$. Equations \eqref{eq_motion_beta_+}--\eqref{eq_motion_beta_-}
	imply that $T^1{}_1=T^2{}_2=T^3{}_3$, \eqref{eq_constraint} is then the usual Friedmann equation, while \eqref{eq_motion_alpha} is the acceleration equation.

 These are the complete evolution equations for the diagonal Bianchi IX case, but,
	as it is well known, all diagonal Bianchi models follow identical equations,
	differing only in the specific form of the potential $U(\beta_+, \beta_-)$.
	The simplest case corresponds to Bianchi I, for which the potential vanishes.
	Thus, in order to obtain the equations for Bianchi I, one can simply set $ U(\beta_+, \beta_-)\to 0$
	in the above equations, which is equivalent to taking the limit $ \rzero \to +\infty $.
	In this sense, the Bianchi I dynamics can be understood as the free dynamics of the
	Bianchi IX model, in particular, and of any other diagonal Bianchi type, in general.
	Consequently, any kinetic-dominated period of Bianchi IX, where the contribution
	of the potential $U(\beta_+,\beta_-)$ is negligible with respect to the remaining (kinetic)
	terms in the equations, can be well approximated by the Bianchi I dynamics.
	We will make use of this fact below in order to analyze	
	the Bianchi IX dynamics in detail.

\section{Main assumptions and gauge choice for an asymptotic analysis}\label{sec.BIX}

The dynamics of the Bianchi IX model is quite involved, and, in particular, it
is extremely difficult to obtain exact analytic solutions. However, as complicated as it may be, its general qualitative behavior
is well known: as proven in Refs.\cite{Lin:1990tq,Lin:1989tv}, provided that the matter satisfies
the dominant energy condition and has a nonnegative average pressure, the
	Bianchi IX model describes a universe with a recollapse.
	Thus $\alpha$ is bounded from above, but not from below, and $\alpha\to-\infty$ corresponds to a singularity where the
	different curvature invariants diverge \cite{Ringstrom:2000mk}. Therefore, the universe is finite in time
	and contains an initial and a final singularity. Moreover, 
	in general, the Bianchi IX dynamics can be well described as a succession
of kinetic-dominated periods, when the potential $U(\beta_+,\beta_-)$ is negligible and thus
the system follows the trajectory given by the Bianchi I dynamics. These periods are interrupted by quick bounces against
the potential walls, which take the system to the next kinetic-dominated period.
Such bounces can be understood as a scattering problem where, given an in-going
state, one would like to obtain the out-going state. In this context,
the relation between the parameters that characterize the out-going and in-going
states is usually called the Kasner transition law, or simply the Kasner map. Its form is well
known for vacuum, and the main goal of this paper is to compute it under the presence
of matter, while the system tends toward the singularity at $\alpha\to-\infty$.

In order to obtain the Kasner map, we need to perform certain approximations.
More precisely, in Subsec.~\ref{sec.assumptions} we impose two well-motivated
assumptions:
$(i)$ the Bianchi IX potential is approximated by a pure exponential, and $(ii)$
matter can be described as a barotropic perfect fluid with a linear equation of state.
Under such approximations, in Sec.~\ref{sec.vacuum}, we obtain the exact solution for vacuum
(as well as for the case
with the perfect fluid being stiff matter), and its corresponding Kasner map.
The general matter content is then considered in Sec.~\ref{sec.generalkasner}, where we linearize the equations
around the vacuum solution, assuming that the matter contribution is small or, equivalently,
that the system is near the singularity. In this way, we can provide the Kasner map
for finite times, when matter begins to matter. Finally, in Sec.~\ref{sec:curvatureBIX}, making use of the obtained
approximate solution, we discuss the behavior of the curvature near the singularity.

\subsection{Main assumptions}\label{sec.assumptions}

\begin{figure}[t]
	\centering
	  \begin{minipage}[t]{0.47\textwidth}
	\includegraphics[width=\linewidth]{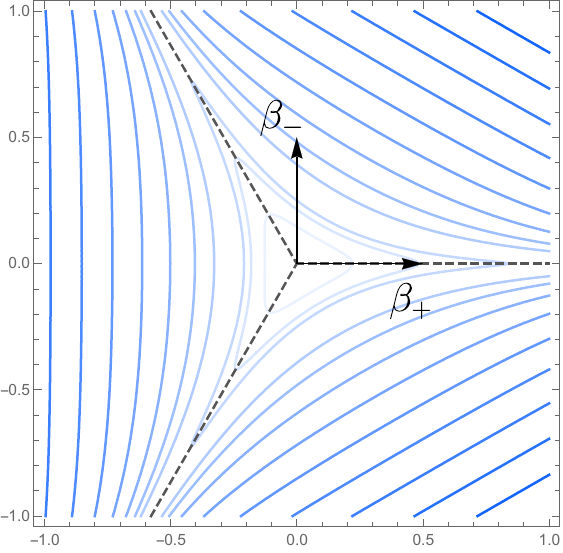}
	\caption{Equipotential plot of the Bianchi IX potential $U(\beta_+,\beta_-)$ given in Eq.~\eqref{potential_expanded}. The solid blue curves are the equipotential lines, with the brightness denoting the value of the potential (the brighter the line the higher the values of the potential). The three dashed black lines are the symmetry axes of the potential, that also correspond to the local minima. Specifically, along these lines the potential takes negative values, ranging from the global minimum at the origin, $U(\beta_+,\beta_-)=-1/2$, to $U(\beta_+,\beta_-)\to 0$ as the shape parameters tend to infinity.}
	\label{fig:equipotentialframes}
	    \end{minipage}
\hfill
\begin{minipage}[t]{0.47\textwidth}
	\includegraphics[width=\linewidth]{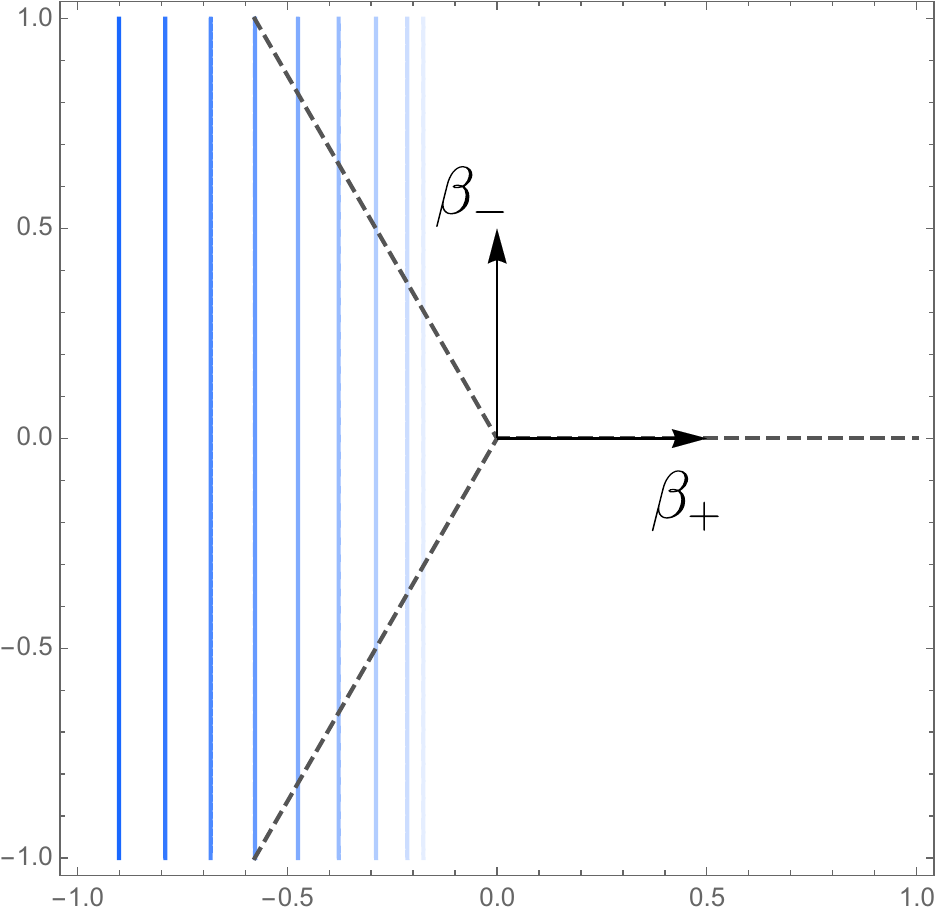}
	\caption{The solid blue lines correspond to the equipotential lines of the function $e^{-8\beta_+}/6$, while the dashed black lines represent the three symmetry axes
	of the Bianchi IX potential $U(\beta_+,\beta_-)$.
		By comparing this with the equipotential plot depicted in Fig.~\ref{fig:equipotentialframes},
	it is clear that $e^{-8\beta_+}/6$ can be considered as
	a good approximation to $U(\beta_+,\beta_-)$ for points in the left wedge
	and located far from the symmetry semiaxes.}
	\label{fig:equipotentialbianchiii2}
	    \end{minipage}
\end{figure}

On the one hand, in order to see how can one approximate the Bianchi IX potential,
let us display its equipotential plot in Fig.~\ref{fig:equipotentialframes}, and
expand its definition \eqref{potential_Bianchi_IX} to write it as a linear combination of exponential terms,
\begin{align}\label{potential_expanded}
	U(\beta_+,\beta_-)=\frac{1}{6}\left(
	e^{-8\beta_+}+e^{4\beta_++4\sqrt{3}\beta_-}+e^{4\beta_+-4\sqrt{3}\beta_-}-2e^{4\beta_+}-2e^{-2\beta_++2\sqrt{3}\beta_-}-2e^{-2\beta_+-2\sqrt{3}\beta_-}
	\right).
\end{align}
As it can be seen, $U(\beta_+,\beta_-)$ has a 3-fold rotational symmetry with respect to the origin. This defines the three symmetry semi-axes
$\{\beta_-=0,\beta_+>0\}$, $\{\beta_-=\sqrt{3}\beta_+, \beta_+<0\}$, and $\{\beta_-=-\sqrt{3}\beta_+, \beta_+<0\}$, which
divide the $(\beta_+,\beta_-)$ plane into three different wedges.
At each of these wedges,
one of the three exponential terms $\{e^{-8\beta_+},e^{4\beta_++4\sqrt{3}\beta_-},e^{4\beta_+-4\sqrt{3}\beta_-}\}$
is dominant in the expression \eqref{potential_expanded} with respect
to the other five.
Without loss of generality, we will assume that the bounce against the potential wall
happens in the left wedge on Fig.~\ref{fig:equipotentialframes},
and thus approximate the potential as $U(\beta_+,\beta_-)\approx e^{-8\beta_+}/6$,
which is the dominant term there. In fact, such form
corresponds to the potential of the Bianchi II model. However,
it is important to point out that this is only a good approximation far away from the symmetry semiaxes, as can be seen in Fig.~\ref{fig:equipotentialbianchiii2},
where the equipotential plot of $e^{-8\beta_+}/6$ is shown.
For bounces taking place in any of the other wedges,
one simply needs to apply a clockwise or an anticlockwise $2\pi/3$
rotation.

On the other hand, we will assume that matter can be described as
a perfect fluid with a linear equation of state. That is, given the velocity
of the fluid $u_\mu$, normalized as $u^\mu u_\mu=-1$, the
energy-momentum tensor reads,
\begin{equation}\label{energy-momentum-tensor}
	T_{\mu\nu}=(\rho+p)u_\mu u_\nu+p g_{\mu\nu},
	\end{equation}
with the pressure $p$ and the energy density $\rho$ being related by
$p=\omega\rho$ with a constant $\omega$. For such fluid, the absence of energy flux \eqref{matterflux}, assuming
$\omega\neq -1$ and $\rho\neq 0$, implies that $u_i=0$, and thus $u_0=N$.\footnote{The value $\omega=-1$ corresponds to a cosmological constant. In that case,
the energy-momentum tensor reads $T_{\mu\nu}=-\rho g_{\mu\nu}$, and it does
involve any velocity vector. The equations \eqref{matterflux} are automatically obeyed,
and the continuity equation leads to $\rho'=0$, and thus $\rho=\rho_0$, which is
included in the general result \eqref{rho_evolution} for $\omega=-1$.}
In addition,
the conservation of the energy-momentum tensor, $\nabla_\mu T^{\mu}{}_{\nu}=0$, provides the continuity equation
	\begin{equation}
	\label{eq_rho_evol}
	{\rho}^\prime+3(\rho+p){\alpha}^\prime=0,
	\end{equation}
which, considering the equation of state $p=\omega\rho$,
is straightforward to solve,
	\begin{equation}
	\label{rho_evolution}
	\rho=\rho_{0} e^{-3(1+\omega)\alpha},
	\end{equation}
	with constant $\rho_{0}$. Note that, even if there may be several
matter components in the model, in general, their decay rate will be different, and each evolutionary
epoch of the universe will be dominated by different components. 
Since we will be interested
in the region close to the singularity $(\alpha\to-\infty)$, our assumption means that we restrict our study to the most relevant matter component in such regime, and we simply disregard the remaining. 
Moreover, we will require that this fluid obeys the dominant
	energy condition, which implies that $\rho$ is nonnegative, that is, $\rho_{0}\geq 0$, and the barotropic index
	$\omega$ is restricted to the interval $-1\leq \omega\leq 1$. The case $\omega=1$
	corresponds to stiff matter and, as will be commented below, it represents a special case, since
	it is the only matter that scales at the same rate as the anisotropies.

In summary, to study the transition between two subsequent kinetic-dominated
periods, we will impose the following two assumptions:
\begin{enumerate}[label=(\roman*)]
	\item The Bianchi IX potential will be approximated by a pure exponential as $U(\beta_+,\beta_-)\approx e^{-8\beta_+}/6$, and thus the bounce
	against the potential wall will take place in the left wedge of Fig.~\ref{fig:equipotentialframes}.
	\item Matter is described by a perfect fluid with a linear equation
	of state $p=\omega\rho$ that follows
	the evolution \eqref{rho_evolution}, with $\rho_0\geq 0$ and $\omega\in[-1,1]$.
\end{enumerate}

\subsection{Gauge fixing}

Before moving to analyze the equations of motion,
at this point we will fix the time gauge by choosing the lapse as $N=e^{3\alpha}$.
For definiteness, we will use $\tau=t$ to refer to the
time in this gauge, and derivatives with respect to $\tau$ will be denoted
by a dot. In this way, the equations of motion \eqref{eq_constraint}--\eqref{eq_motion_beta_-} read,
	\begin{align}
	\label{constraint_lapse}
	0&=\kappa\rho-3 e^{-6\alpha}\left({\dot{\alpha}}^2-{\dot{\beta}}_+^2-{\dot{\beta}}_-^2\right)+\frac{6}{ \rzero^2}e^{-2 \alpha}U(\beta_+,\beta_-),
	\\
	\label{eq_motion_alpha_lapse}
	\ddot{\alpha}&=\frac{\kappa}{2}e^{6\alpha}\left(\rho -p\right)+\frac{4}{\rzero^2}e^{4 \alpha}\,U(\beta_+,\beta_-),
	\\
	\label{eq_motion_beta_+_lapse}
	\ddot{\beta}_+&= -\frac{e^{4 \alpha}}{\rzero^2}\,\frac{\partial U(\beta_+,\beta_-)}{\partial\beta_+},\\
	\label{eq_motion_beta_-_lapse}
	\ddot{\beta}_-&= -\frac{e^{4 \alpha}}{\rzero^2}\,\frac{\partial U(\beta_+,\beta_-)}{\partial\beta_-}.
	\end{align}
In this form, it is explicit that the evolution equations for the shape parameters are equivalent to those for a point particle moving
	on the plane $(\beta_+,\beta_-)$ under the influence of the potential $e^{4 \alpha} U(\beta_+,\beta_-)/\rzero^2$.
	{Toward the singularity $\alpha\to-\infty$, the factor $e^{4\alpha}$ makes the value of the potential at each point $(\beta_+,\beta_-)$ to diminish,
		and thus the potential walls are effectively pushed back from the origin of the plane $(\beta_+,\beta_-)$.
		Consequently, there is a region where the potential becomes negligible, which implies that the system moves freely, i.e., following a Bianchi I dynamics, until colliding with a potential wall.
		Hence, in this limit, it becomes particularly clear that the Bianchi IX dynamics can be approximated as a succession of Bianchi I periods.
	}

Imposing now the approximate form of the potential $U(\beta_+,\beta_-)\approx e^{-8 \beta_+}/6$ commented above, together with the linear equation of state and the evolution of the density \eqref{rho_evolution}, the evolution equations \eqref{eq_motion_alpha_lapse}--\eqref{eq_motion_beta_-_lapse} read,
\begin{align}\label{eq_motion_alpha_II}
\ddot{\alpha}&=\frac{\kappa}{2}(1-\omega)\rho_0e^{3(1-\omega)\alpha}+\frac{2}{3\rzero^2}e^{4\alpha-8\beta_+},
\\
\label{eq_motion_beta_+_II}
\ddot{\beta}_+&=\frac{4}{3\rzero^2}e^{4\alpha-8\beta_+},
\\
\label{eq_motion_beta_-_II}
\ddot{\beta}_-&=0,
\end{align}
and the constraint \eqref{constraint_lapse} takes the simpler form,
\begin{align}\label{eq_constraint_II}
\kappa\rho_0e^{-3(1+\omega)\alpha}
-3e^{-6\alpha}\left(
\dot{\alpha}^2-\dot{\beta}_+^2-\dot{\beta}_-^2
\right)+\frac{1}{\rzero^2}e^{-2(\alpha+4\beta_+)}=0.
\end{align}
The Eq.~\eqref{eq_motion_beta_-_II} for the shape parameter $\beta_-$ can be exactly solved,
\begin{equation}
\label{sol_beta_-_exact}
 \beta_-=k_-+p_- \tau,
\end{equation}
where $p_-$ and $k_-$ are two integration constants. Nonetheless, the remaining equations for $\alpha$ and $\beta_+$ are nontrivial. At this point, it is convenient to introduce the variables $x:=2\alpha-\beta_+$ to replace $\alpha$.
In this way, the second-order equations \eqref{eq_motion_alpha_II}--\eqref{eq_motion_beta_+_II}
can be rewritten in the form,
\begin{align}\label{xdotdot}
\ddot{x}&=
\kappa (1-\omega)\rho_0e^{\frac{3}{2}(1-\omega)(x+\beta_+)},
\\
\label{betadotdot}
	\ddot{\beta}_+&=\frac{4}{3\rzero^2}e^{2x-6\beta_+},
 \end{align}
 where only the evolution of $x$ is explicitly coupled to the matter field, 
while
the constraint \eqref{eq_constraint_II} reads
 \begin{align}\label{constraint_II_x_y}
 \kappa\rho_0e^{-\frac{3}{2}(1+\omega)(x+\beta_+)}
 -3e^{-3(x+\beta_+)}\left[\frac{1}{4}\left(\dot{x}+\dot{\beta}_+\right)^2-\dot{\beta}_+^2-p_-^2
 \right]+\frac{1}{\rzero^2}e^{-x-9\beta_+}=0.
 \end{align}
In summary, under the assumptions $(i)$ and $(ii)$ above, the analysis of the
Bianchi IX dynamics is reduced to obtaining the solution to the system of
equations \eqref{xdotdot}--\eqref{constraint_II_x_y} for $x$ and $\beta_+$.
In general this is still a nontrivial task, although there are two particular cases
that make the right-hand side of \eqref{xdotdot} to be vanishing, and thus
greatly simplify the resolution
of the system: $\omega=1$
and $\rho_0=0$. The former corresponds to stiff matter, and the latter to vacuum.
In fact, 
from equations \eqref{xdotdot}--\eqref{betadotdot}, both vacuum and stiff-matter content yield
formally the same solution for both $x$ and $\beta_+$. And, although
the form of their corresponding constraint \eqref{constraint_II_x_y} differs,
the stiff-matter contribution $\frac{\kappa}{3}\rho_{0{\rm stiff}}$ can be absorbed in the constant
$p_-$. This is what we will do in the next section in order to treat both cases
in the same footing.

Finally, from Eq.~\eqref{xdotdot} we see that, for $\omega\neq 1$,
the vacuum solution is also recovered when the exponential term $e^{\frac{3}{2}(1-\omega)(x+\beta_+)}$ tends to zero, regardless the (finite) value of $\rho_0$. This happens precisely when $(x+\beta_+)=2\alpha\to -\infty$, that is, as the system approaches the singularity. Therefore, toward this limit, the dynamics of the model with any matter content (with $\omega\neq 1$ and $\rho_0\neq 0$) tends to the vacuum solution. This is, precisely, the famous ``matter does not matter'' statement mentioned in the introduction, which, however, it is obeyed only in the exact $\alpha\to-\infty$ limit. For finite, though large
values of $|\alpha|$, matter effects do affect the evolution of the model.

\section{Kasner map for vacuum and stiff-matter content}\label{sec.vacuum}

Now we will study the transition law that connects two vacuum Bianchi I epochs
for the particular cases of vacuum (Subsec.~\ref{sec.vacuum.kasner}) and stiff-matter content (Subsec.~\ref{sec.kasner.stiff}).
In fact, the results for a stiff-matter content can be easily derived from the corresponding
vacuum results by simply performing the replacement $p_-^2\to p_-^2+\frac{\kappa}{3}\rho_{0{\rm stiff}}$. However the stiff-matter content presents some interesting
features, and thus we will provide its explicit Kasner map and
discuss it in detail in Subsec.~\ref{sec.kasner.stiff}.

	Since we will need it in the analysis below,
	let us briefly present here the exact solution
	for the Bianchi I model.
	As commented above,
	the equations of motion for the Bianchi I model can be obtained from \eqref{constraint_lapse}--\eqref{eq_motion_beta_-_lapse} by simply setting $U(\beta_+,\beta_-)=0$. Therefore, for any perfect fluid, the shape parameters follow a linear evolution in the $\tau$ time,
	\begin{align}\label{sol_beta_+_Kasner}
	\beta_+&=k_++p_+ \tau,\\
	\label{sol_beta_-_Kasner}
	\beta_-&=k_-+p_- \tau,
	\end{align}
	with constants $k_+,k_-,p_+,$ and $p_-$. For an exact Bianchi I geometry,
	the constants $k_\pm$ are pure gauge and can be absorbed in a redefinition
	of the coordinates, while $p_\pm$ are the canonical momenta of $\beta_\pm$,
	and completely encode the spatial anisotropy.
	In particular, the isotropic case corresponds to $p_+=0=p_-$. In addition,
	for both vacuum and stiff-matter content, i.e. with $\rho_0=\rho_{0{\rm stiff}}\geq 0$, it is straightforward to obtain
	$\alpha$ from the constraint \eqref{constraint_lapse},
	\begin{equation}
		\label{sol_a_Kasner_vacuum}
		\alpha=\pm\left(p_+^2+p_-^2+\frac{\kappa}{3}\rho_{0{\rm stiff}}\right)^{\!1/2}\tau+c_\alpha,
	\end{equation}
with a constant $c_\alpha$, while the sign $\pm$ characterizes the expanding
and contracting branches of the solution. For vacuum, $\rho_{0{\rm stiff}}=0$, and this reduces to the well-known Kasner solution \cite{Kasner:1921zz}.

\subsection{Kasner map for vacuum}\label{sec.vacuum.kasner}

Imposing $\rho_0=0$ in the Bianchi IX evolution for $x$ \eqref{xdotdot}, one immediately gets
\begin{equation}\label{sol_x_vacuum_stiff}
 x=C_1+C_2 \tau,
\end{equation}
with integration constants $C_1$ and $C_2$. Using this,
and introducing the variable $q:=3\beta_+-x$,
the constraint \eqref{constraint_II_x_y} can be written as,
\begin{equation}\label{qeq}
\frac{1}{2}\dot{q}^2+\Phi(q)=0,
\end{equation}
where we have defined
\begin{equation}
 \Phi(q):=-2\left(\dot{x}^2-3 p_-^2-\frac{e^{-2q}}{\rzero^2} \right).
\end{equation}
Note that here, for convenience, we have explicitly left $\dot{x}$ indicated,
though it is a constant and thus $\Phi$ only depends on $q$.
From this expression it is clear that, since $e^{-2 q}$ is positive definite, 
\begin{equation}\label{condition}
0<\dot{x}^2-3 p_-^2,
\end{equation}
or, equivalently,
\begin{equation}\label{condition2}
0<(2\dot{\alpha}-\dot{\beta_+})^2-3 p_-^2,
\end{equation}
is a necessary condition for the existence of the solution. In addition,
\eqref{qeq} can be interpreted as the conservation of the total (null) energy
for a particle $q=q(\tau)$ moving on the potential $\Phi(q)$.
This potential is continuous and bounded from below, with $\Phi(q)\to +\infty$ as $q\to-\infty$
and $\Phi(q)\to -2 (\dot{x}^2-3 p_-^2)$ as $q\to+\infty$. Thus $\Phi(q)$ has exactly
one root $q=q_0$, with $q_0:=-\frac{1}{2}\ln[\rzero^2 (\dot{x}^2-3p_-^2)]$, where
a bounce happens. That is, a given trajectory that begins at $q\to+\infty$ with $\tau\to -\infty$
and a velocity $\dot{q}=-2(\dot{x}^2-3p_-^2)^{1/2}$,
then evolves toward lower values of $q$ until it reaches $q_0$ with $\dot{q}=0$.
Here a bounce happens, and then it evolves back toward $q\to+\infty$ as $\tau\to+\infty$.

In fact, one can explicitly obtain the general solution of \eqref{qeq} by direct integration,
\begin{equation}\label{solq}
q=-\ln\left[\rzero C_3
\sech\left(2C_3 \tau
\right)\right],
\end{equation}
where we have defined $C_3:=(C_2^2-3p_-^2)^{1/2}>0$, and imposed the initial
condition $q(0)=q_0$. This condition fixes the origin of time at the bounce
and makes the function $q(\tau)$ symmetric, i.e., $q(\tau)=q(-\tau)$.

In order to obtain the evolution of our original variables $(\alpha, \beta_+)$,
one can simply invert the definitions of $x$ and $q$ to get,
\begin{align}\label{sol_alpha_vacuum_stiff}
	\alpha&=\frac{1}{6}(4x+q) =\frac{2}{3}(C_1+C_2\tau)
	-\frac{1}{6}\ln\left[\rzero C_3
	\sech\left(2C_3\tau
	\right)\right], \\\label{sol_beta_+_vacuum_stiff}
	\beta_+ &=\frac{1}{3}(x+q)\,\,\,=\frac{1}{3}(C_1+C_2\tau)
-\frac{1}{3}\ln\left[\rzero C_3
\sech\left(2C_3\tau
\right)\right].
\end{align}
These solutions encompass the entire dynamics before, during, and after the bounce against the potential wall.
Now, for concreteness and without loss of generality,
we will consider that the singularity $\alpha\to -\infty$ is located at $\tau\to-\infty$,
and thus we will evolve the system backward in time. Note that, in particular,
this implies that $\dot{\alpha}>0$, which, considering \eqref{condition} and \eqref{sol_alpha_vacuum_stiff}, leads to the condition
$C_2>\sqrt{3} |p_-|$. In this way, $\alpha=\alpha(\tau)$ is monotonic,
going from $\alpha\to-\infty$ at $\tau\to-\infty$, and $\alpha\to+\infty$ at $\tau\to+\infty$. However,
$\beta_+(\tau)$ is not monotonic in general; while $\beta\to+\infty$ as $\tau\to+\infty$,
depending on the values of the different parameters, it might
go either to $-\infty$ or to $+\infty$ as $\tau\to-\infty$.
Since the bounce occurs at $\tau=0$,
we will name $\tau>0$ and $\tau<0$ the pre-bounce and post-bounce periods, respectively.

In the limits where the potential term $e^{-2 q}/\rzero^2$ is negligible, i.e., at $q\to +\infty$, and hence $\tau\to\pm\infty$, the evolution of the variables \eqref{sol_beta_-_exact}, \eqref{sol_alpha_vacuum_stiff}, and \eqref{sol_beta_+_vacuum_stiff} will
tend to their corresponding form  in vacuum Bianchi I
\eqref{sol_beta_+_Kasner}--\eqref{sol_a_Kasner_vacuum}, that is,
	\begin{equation}
\label{sol_Kasner_vacuum}
\begin{aligned}
\alpha&=P\tau+c_\alpha,
\\
\beta_+&=p_+\tau+k_+,
\\
\beta_-&=p_-\tau+k_-,
\end{aligned}
\end{equation}
with $P:=(p_+^2+p_-^2)^{1/2}$. Note that here we have imposed $\dot{\alpha}=P$ to be positive
in order to match the choice taken above for the general solution \eqref{sol_alpha_vacuum_stiff}
so that the singularity is located at $\tau\to -\infty$.
Therefore, each kinetic-dominated period\
is completely characterized by the five parameters $\{p_+, p_-,k_+,k_-,c_\alpha\}$.\footnote{\label{footnote}Note that, among these five parameters, only four are free,
since in \eqref{sol_Kasner_vacuum} the time gauge is
not completely fixed
(the Hamiltonian constraint is solved, but there is still the freedom
to choose the origin of time).
It is more convenient to work with these five parameters, since
we will have to relate \eqref{sol_Kasner_vacuum}
to the asymptotics of \eqref{sol_alpha_vacuum_stiff}--\eqref{sol_beta_+_vacuum_stiff}, where the gauge is completely fixed and thus certain gauge choices could be inconsistent.
In any case, below we will provide the explicit expression that constraints these five parameters.}
The parameters corresponding to the pre-bounce phase will be denoted with an overline,
and those corresponding to the post-bounce phase with a tilde.
In this way, the Kasner map will provide the post-bounce state
$\{\widetilde p_+,\widetilde p_-,\widetilde k_+,\widetilde k_-,\widetilde c_\alpha\}$ in terms of the pre-bounce state
$\{\overline p_+,\overline  p_-,\overline k_+,\overline k_-,\overline c_\alpha\}$.

In particular, the map for the parameters characterizing the variable $\beta_-$
is trivial: since its exact solution \eqref{sol_beta_-_exact}, which is valid
for the whole evolution, coincides with its form \eqref{sol_Kasner_vacuum} during the kinetic-dominated periods,
we will simply have that $\widetilde p_-=\overline p_-$ and $\widetilde k_-=\overline k_-$. We now proceed to construct the Kasner map for the remaining
parameters associated to the variables $\alpha$ and $\beta_+$.

\begin{itemize}
	\item \textbf{Before the bounce} ($\tau\to +\infty$):

On the one hand, in the limit $\tau\to +\infty$ the
solutions 
\eqref{sol_alpha_vacuum_stiff} and
\eqref{sol_beta_+_vacuum_stiff} tend to
	\begin{align}\label{sol_alpha_vacuum_stiff_ini}
	\alpha&=
	\frac{\tau}{3}(2C_2+C_3)
	+\frac{1}{6}\left[4C_1-\ln(2\rzero C_3)\right],
	\\
	\label{sol_beta_+_vacuum_stiff_ini}
	\beta_+&=\frac{\tau}{3}(C_2+2C_3)
	+\frac{1}{3}\left[C_1-\ln(2\rzero C_3)\right].
 \end{align}
Comparing these results with the Bianchi I dynamics \eqref{sol_Kasner_vacuum}, parametrized by the constants $\{\overline p_+, \overline  k_+,\overline  c_\alpha\}$,
we conclude that the pre-bounce state is given by
	\begin{align}\label{Kasner_param_ini}
		\begin{aligned}
	\overline p_+&=\frac{1}{3}(C_2+2C_3),
	\\
	\overline k_+&=\frac{1}{3}\left[C_1-\ln(2\rzero C_3)\right],
	\\
	\overline 
	c_\alpha&=\frac{1}{6}\left[4C_1-\ln(2\rzero C_3)\right].
		\end{aligned}
	\end{align}
From here, we note that $C_3=2\overline p_+-\overline P$, where $\overline P:=(\overline p_+^2+\overline p_-^2)^{1/2}$, and, since $C_3$ is positive, we conclude that all
the pre-bounce states obey $\overline p_+>|\overline p_-|/\sqrt{3}$. Therefore, in this regime, the only Bianchi I trajectories that exist are those that satisfy this condition.

	\item \textbf{After the bounce} ($\tau\to -\infty$):

	On the other hand, in the limit $\tau\to -\infty$, the solutions
\eqref{sol_alpha_vacuum_stiff} and \eqref{sol_beta_+_vacuum_stiff} tend to
	\begin{align}\label{sol_alpha_vacuum_stiff_fin}
	\alpha&=
	\frac{\tau}{3}(2C_2-C_3)
	+\frac{1}{6}\left[4C_1-\ln(2\rzero C_3)\right],
	\\
	\label{sol_beta_+_vacuum_stiff_fin}
	\beta_+&=\frac{\tau}{3}(C_2-2C_3)
	+\frac{1}{3}\left[C_1-\ln(2\rzero C_3)\right].
	\end{align}
	
	Again, comparing these results with the dynamics \eqref{sol_Kasner_vacuum} in Bianchi I, now parametrized by the constants $\{\widetilde p_+,\widetilde k_+,\widetilde c_\alpha\}$, one concludes that in the post-bounce state:
	\begin{align}\label{Kasner_param_fin}
	\begin{aligned}
	\widetilde p_+&=\frac{1}{3}(C_2-2C_3),
	\\
	\widetilde k_+&=\frac{1}{3}\left[C_1-\ln(2\rzero C_3)\right],
	\\
	\widetilde 
	c_\alpha&=\frac{1}{6}\left[4C_1-\ln(2\rzero C_3)\right].
	\end{aligned}
	\end{align}
In this case, we observe that $C_3=\widetilde P-2\widetilde p_+$, where $\widetilde P:=(\widetilde p_+^2+\widetilde p_-^2)^{1/2}$. By the same reasoning as before, since this quantity is positive by definition, we conclude that all post-bounce states obey $\widetilde p_+<|\widetilde p_-|/\sqrt{3}$, contrary to the pre-bounce state. Therefore, after the bounce, in the asymptotic regime closer to the singularity, the only Bianchi I trajectories that exist are those that satisfy this condition.
\end{itemize}
Then, in order to obtain the transition law, one can simply
solve the constants $(C_1,C_2)$ in terms of the initial parameters
$\{\overline p_+,\overline k_+,\overline c_\alpha\}$ from \eqref{Kasner_param_ini}
and replace them in \eqref{Kasner_param_fin}.
This straightforward computation leads to the vacuum Kasner map,
\begin{align}\label{transition_stiff_vacuum}
\begin{aligned}
\widetilde{p}_+&=\frac{1}{3}\left(
4\overline P-5\overline p_+
\right),
	\\
\widetilde{k}_+&=\overline k_+,
	\\
\widetilde{c}_\alpha&=\overline c_\alpha,
	\\
\widetilde{p}_-&=\overline p_-,
	\\
\widetilde{k}_-&=\overline k_-,
\end{aligned}
\end{align}
where $\overline P:=(\overline p_+^2+\overline p_-^2)$, and we have also included
the trivial map for the parameters associated to $\beta_-$. That is,
all the parameters are conserved through the bounce, except $p_+$.
Note that, in this simple case,
$c_\alpha$, $k_+$, and $k_-$ are constants of the full dynamics, as they are conserved through the bounce,
and thus they can be reabsorbed in the coordinates by a global (time-independent)
coordinate transformation. However, as will be shown below, this will not be the case when considering
a general barotropic fluid.
Also, as commented in the footnote \ref{footnote}, these constants are not independent. From
the asymptotic forms \eqref{sol_alpha_vacuum_stiff_ini}--\eqref{sol_beta_+_vacuum_stiff_ini} and \eqref{sol_alpha_vacuum_stiff_fin}--\eqref{sol_beta_+_vacuum_stiff_fin}, it is easy to check
that they obey the constraint
\begin{equation}\label{rel_k_+_alpha}
  k_+=\frac{c_\alpha}{2}-\frac{1}{4}\ln[2 \rzero|P-2p_+|].
\end{equation}

Let us now analyze the physical implications of the map \eqref{transition_stiff_vacuum}.
During the kinetic-dominated periods, the velocity vector
{$\vec{v}:=(-\dot{\beta}_+,-\dot{\beta}_-)=(-p_+,-p_-)$,
defined with the negative sign as we are considering the backward evolution in time,
is constant and its Euclidean norm is $P$. Thus, the system follows a straight line in the plane of anisotropies $(\beta_+,\beta_-)$. However,
the bounce modifies the direction and norm of such vector in the way predicted
by \eqref{transition_stiff_vacuum}.
In this respect, it
is convenient to define $\theta\in[0,2\pi)$ as the
angle between the velocity vector
and the $\beta_+$ axis, that is, $\sin\theta:=-p_-/P$ and
$\cos\theta:=-p_+/P$. Consequently, instead of $(p_+,p_-)$,
one can equivalently choose the magnitude $P$ and the
polar angle $\theta$ to describe the pre-bounce
$(\overline P, \overline\theta)$ and post-bounce
$(\widetilde P, \widetilde\theta)$
velocity vectors, and write their corresponding Kasner map, namely,
\begin{align}
\label{trans_P}
 \widetilde P =&\frac{\overline P}{3}\left(5+4\cos\overline\theta\right),\\\label{trans_thetacos}
 \cos\widetilde\theta =&
 -\frac{4+5\cos\overline\theta}{5+4\cos\overline\theta},\\ \label{trans_theta}
 \sin\widetilde{\theta}=&\frac{3\sin\overline\theta}{5+4\cos\overline\theta}.
\end{align}
 Let us recall that, since the pre-bounce trajectories must satisfy the condition $\overline p_+>|\overline p_-|/\sqrt{3}$, the pre-bounce angle $\overline\theta$ is limited to the range $\overline\theta\in(2\pi/3,4\pi/3)$. Therefore, the above transition applies exclusively to angles within this range. Additionally, post-bounce trajectories are characterized by the complementary property, $\widetilde p_+>|\widetilde p_-|/\sqrt{3}$, meaning that the post-bounce angle $\widetilde\theta$ lies in the range $[0,2\pi/3)\cup(4\pi/3,2\pi)$.
 Note that $\overline\theta=2\pi/3$ and $\overline\theta=4\pi/3$ are not included in the range of pre-bounce angles, but, 
if one evaluates the map \eqref{trans_P}--\eqref{trans_theta} on these angles, it is easy to see that it is the identity. Therefore, one can extend the map to the closed interval
$\overline\theta\in[2\pi/3,4\pi/3]$ by continuity, which corresponds to the range $\overline p_+\in[\overline P/2,\overline P]$ for the pre-bounce $\overline p_+$.

Furthermore, the norm of the velocity generically decreases during
the bounce and its post-bounce value can vary within the range $\widetilde P\in[\overline P/3,\overline P]$. Specifically, according to \eqref{transition_stiff_vacuum}, while the vertical
component of the velocity $p_-$ remains constant, the reduction in $P$ is attributed
to a change in the horizontal component $p_+$, with post-bounce values
in the range $\widetilde p_+\in[-\widetilde P,\widetilde P/2]$. More precisely, one can define two types of bounces:
	
	\begin{enumerate}[label=(\alph*)]
		\item \textit{Backward scattering}: the sign of the horizontal component of the velocity changes, resulting in $\widetilde{p}_+ < 0$. This occurs for pre-bounce angles $\overline{\theta} \in (\arccos(-4/5), 2\pi - \arccos(-4/5))$, and the post-bounce norm $\widetilde{P}$ lies in the range $\widetilde{P} \in [\overline{P}/3, 3\overline{P}/5)$.
		\item \textit{Forward scattering}: the sign of the horizontal component of the velocity does not change, hence $\widetilde{p}_+ > 0$. This scenario corresponds to pre-bounce angles $\overline{\theta} \in (2\pi/3, \arccos(-4/5)) \cup (2\pi - \arccos(-4/5), 4\pi/3)$, and the post-bounce norm $\widetilde{P}$ lies within the range $\widetilde{P} \in (3\overline{P}/5, \overline{P}]$.
	\end{enumerate}
	Therefore, backward scatterings result in a more significant decrease in the norm of the velocity compared to forward scatterings. In particular, the maximum change
	in both the angle and norm of the velocity vector occurs for a head-on collision, where the trajectory aligns parallel to the $\beta_+$ axis, corresponding to the pre-bounce state $\overline{\theta} = \pi$ ($\overline{p}_+ = \overline{P}$). After the bounce, this configuration leads to $\widetilde{\theta} = 0$ ($\widetilde{p}_+ = -\widetilde{P}$) and $\widetilde{P}=\overline{P}/3$, which is the lower bound for the post-bounce norm. Conversely, trajectories with $\overline{\theta} = 2\pi/3$ or $\overline{\theta} = 4\pi/3$ (which both correspond to $\overline p_+=\overline P/2$) do not bounce,
	and the map \eqref{transition_stiff_vacuum} is the identity.

To conclude, it is important to note that, in computing this map as an approximation to the full Bianchi IX dynamics,
we have relied on certain assumptions (see Subsection \ref{sec.assumptions}). In particular, we have assumed that the bounce happens against a potential wall located in the left wedge of Fig.~\ref{fig:equipotentialframes}. Therefore, bounces in this region occur when the pre-bounce angle $\overline\theta$ lies in the range $\overline\theta\in(2\pi/3,4\pi/3)$. However, for the Bianchi IX model, since there are potential walls in every other
region of the $(\beta_+,\beta_-)$ plane, trajectories with any other pre-bounce angle are allowed.
Specifically, due to the three-fold rotational symmetry of the potential \eqref{potential_Bianchi_IX}, for $\overline\theta\in(0,2\pi/3)$ the bounce takes place in the upper-right wedge, and for  $\overline\theta\in(4\pi/3,2\pi)$, it occurs in the lower-right one.
The boundary angles $\overline\theta=0,2\pi/3$, and $4\pi/3$ correspond precisely to the symmetry semiaxes of the potential, where the potential tends to zero as the shape parameters approach infinity. In the full Bianchi IX model, not just considering a single exponential term as the potential, trajectories with such angles are allowed approaching the singularity. This happens because, along these directions, all exponential terms in the potential \eqref{potential_expanded} asymptotically tend to zero or to a finite constant. As a result, the dynamics remains kinetic-dominated all along until reaching the singularity without any bounce against the potential walls. Thus, for these specific directions, it can be understood that the system --- viewed as a particle moving on the $(\beta_+,\beta_-)$ plane --- escapes along these axes from the basin formed by the potential.

\subsection{Kasner map for a stiff-matter content}\label{sec.kasner.stiff}

For completeness, let us now explicitly consider the case with a
stiff-matter content. For such case, as commented above,
its corresponding Kasner map can be obtained by performing
the change $p_-^2\to p_-^2+\frac{\kappa}{3}\rho_{0{\rm stiff}}$
in \eqref{transition_stiff_vacuum}, both in the pre- and post-bounce quantities.
In this way, as in the vacuum case,
all the parameters are conserved, except $p_+$, which changes according to
\begin{equation}\label{kasnermap_stiffmatter}
\widetilde p_+=\frac{1}{3}\left(4\overline P\sqrt{1+\frac{\kappa}{3\overline P^2}\rho_{0{\rm stiff}}}-5\overline p_+\right).
\end{equation}
Equivalently, one can write this for the norm and polar angle of the velocity vector:
\begin{align}\label{trans_P_stiff}
\widetilde P &=\frac{\overline P}{3}\left[
\left(5\,\sqrt{1+\frac{\kappa}{3\overline P^2}\rho_{0{\rm stiff}}}+4\cos\overline\theta
\right)	^2-\frac{3\kappa}{\overline P^2}\rho_{0{\rm stiff}}
\right]^{1/2},
\\[6pt]
\label{trans_theta_stiff_1}
\cos\widetilde\theta &=
-\frac{4\,\sqrt{1+\frac{\kappa}{3\overline P^2}\rho_{0{\rm stiff}}}+5\cos\overline\theta}{\left[
	\left(5\,\sqrt{1+\frac{\kappa}{3\overline P^2}\rho_{0{\rm stiff}}}+4\cos\overline\theta
	\right)	^2-\frac{3\kappa}{\overline P^2}\rho_{0{\rm stiff}}
	\right]^{1/2}},
\\
\label{trans_theta_stiff_2}
\sin\widetilde\theta &=
\frac{3\sin\overline\theta}{\left[
	\left(5\,\sqrt{1+\frac{\kappa}{3\overline P^2}\rho_{0{\rm stiff}}}+4\cos\overline\theta
	\right)	^2-\frac{3\kappa}{\overline P^2}\rho_{0{\rm stiff}}
	\right]^{1/2}}.
\end{align}
As can be seen, the matter contribution in these last
equations appears divided by the norm of the
pre-bounce velocity $\overline P^2$. Thus if $\kappa\rho_{0{\rm stiff}}/\overline P^2$ is small,
the matter effects will be negligible. 

First of all, let us analyze the ranges of validity of $\overline \theta$ for this Kasner map. In the vacuum case, we have identified $C_3$, which must be positive for the existence of the solution, in terms of the
pre-bounce parameters as $C_3=2\overline p_+-\overline P$, and then checked its sign.
Thus, by performing the replacement $p_-^2\to p_-^2+\frac{\kappa}{3}\rho_{0{\rm stiff}}$, for stiff matter one obtains the condition
\begin{align}\label{c_3_stiff}
	C_3=2\overline p_+-\sqrt{\overline p_+^2+\overline p_-^2+\frac{\kappa}{3}\rho_{0{\rm stiff}}}>0,
\end{align}
which implies that
\begin{align}\label{cond_matter}
	\overline p_+>\frac{1}{3}\left(
	3\overline p_-^2+\kappa\rho_{0{\rm stiff}}
	\right)^{1/2},
\end{align}
or, equivalently,
\begin{equation}
\kappa\rho_{0{\rm stiff}}<9 \overline P^2-12 \overline{p}_-^2.
\end{equation}
For fixed values of $\overline P$ and $\rho_{0{\rm stiff}}\geq 0$, this condition provides a range
of $\overline p_-$ for which the solution exists.
In particular, it is easy to see that this condition is fulfilled for some $\overline p_-$ if and only if
\begin{equation}\label{max_rho_stiff}
\frac{\kappa\rho_{0{\rm stiff}}}{\overline P^2}<9.
\end{equation}
Consequently, for a fixed value of $\overline P$, there is a maximum allowed density $\rho_{0{\rm stiff}}$.
Equivalently, this can be seen in terms of the pre-bounce angle $\overline\theta$,
for which the condition \eqref{c_3_stiff} reads
\begin{align*}
\cos\overline\theta<-\frac{1}{2}\sqrt{1+\frac{\kappa}{3\overline P^2}\rho_{0{\rm stiff}}}.
\end{align*}
From here, one can define the angle 
\begin{align}\label{theta_max_stiff}
\theta_{\rm stiff}:=\arccos\left[-\frac{1}{2}\left(1+\frac{\kappa}{3\overline P^2}\rho_{0{\rm stiff}}\right)^{1/2}\right],
\end{align}
providing the range of validity for the Kasner map \eqref{trans_P_stiff}--\eqref{trans_theta_stiff_2} for
stiff-matter content as $\overline\theta\in\left(
\theta_{\rm stiff},2\pi-\theta_{\rm stiff}\right)$.
In addition,
according to \eqref{trans_theta_stiff_1}--\eqref{trans_theta_stiff_2},
the post-bounce angle lies within the range $\widetilde\theta\in[0,\theta_{\rm stiff}]\cup[2\pi-\theta_{\rm stiff},2\pi)$, which is broader than in the vacuum case. This is illustrated in Fig.~\ref{fig:postpre_angles}.
In particular, it is important to note that $\theta_{\rm stiff}$ is precisely the angle at which the Kasner map \eqref{trans_P_stiff}--\eqref{trans_theta_stiff_2} becomes the identity. Hence, even if the solution is not defined in the present approximation,
by continuity, we can include $\theta_{\rm stiff}$ in the range of possible values for $\overline\theta$.
Naturally, in the limit $\rho_{0{\rm stiff}}\to 0$, one recovers the vacuum range $\overline\theta\in[2\pi/3,4\pi/3]$. However, as the ratio $\kappa\rho_{0{\rm stiff}}/\overline P^2$ increases, the range
of $\overline\theta$ gradually narrows, until condition \eqref{max_rho_stiff} is not fulfilled,
which implies that the range of $\overline\theta$ is empty and there is no solution anymore.

Once that the range of validity is established, let us study the effects of matter on the Kasner map, as compared to the vacuum scenario. First, from \eqref{kasnermap_stiffmatter}, we observe that
the energy density $\rho_{0{\rm stiff}}$ increases the value of the post-bounce velocity component
$\widetilde p_+$. Consequently, since the component $p_-$ is conserved, the post-bounce angle $\widetilde{\theta}$ is closer to $\pi$, that is, the post-bounce velocity vector forms a sharper
angle with the negative $\beta_+$ axis as compared to its vacuum counterpart.
This implies that, for the same pre-bounce $\overline\theta$, the deflection angle $\Delta\theta:=|\widetilde\theta-\overline\theta|$ is always smaller with matter (see Fig.~\ref{fig:post-bounce}).
This can be understood as the collision with the wall being more
defocusing than in the vacuum case. Therefore, the stiff-matter contribution effectively increases
the convex curvature of the wall. This could have significant implications for the chaotic behavior of the system, as convex and thus defocusing walls are a prerequisite for chaos to manifest in this kind of bouncing systems \cite{Sinai:1970}.

	\begin{figure}
	\centering
	\begin{subfigure}[b]{0.45\textwidth}
		\centering
		\includegraphics[width=\linewidth]{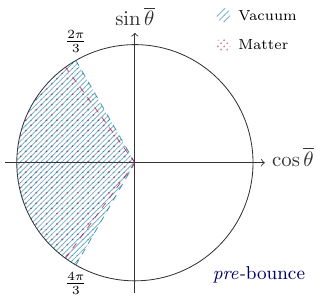}
	\end{subfigure}
	\hfill
	\begin{subfigure}[b]{0.45\textwidth}
		\centering
		\includegraphics[width=\textwidth]{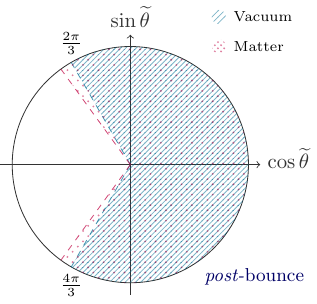}
	\end{subfigure}
	\caption{Comparison of the pre-bounce and post-bounce angles of the velocity vector in vacuum and with matter. The left plot depicts the possible values of the angle before the bounce, denoted as $\overline{\theta}$, while the right plot shows the values of the angle after the bounce, denoted as $\widetilde{\theta}$. It can be observed that, in the presence of matter, the range of $\overline{\theta}$ decreases, whereas the range of $\widetilde{\theta}$ increases. This, viewing the bounce as a scattering problem for a beam, can be understood as the system being more defocusing with matter as compared to vacuum, since a narrower beam scatters to create a wider beam. It is important to note that in both scenarios --- vacuum and with matter --- the pre-bounce and post-bounce ranges are complementary and together span the full range $[0, 2\pi)$. Therefore, every angle within the range $[0, 2\pi)$ is either a pre-bounce or a post-bounce angle, but not both simultaneously; except for the limiting angles (for vacuum, $2\pi/3$ and $4\pi/3$), where the Kasner map is the identity, as there is no bouncing.
	}
	\label{fig:postpre_angles}
\end{figure}

However, the effect of the bounce in the norm of the velocity vector varies. In fact, since $\widetilde p_-$ is the same both for vacuum and stiff matter,
it is determined by the sign of $\widetilde{p}_+$.
This component is always bigger than in vacuum, according to \eqref{kasnermap_stiffmatter}, so its absolute value is greater than the vacuum one only when $\widetilde{p}_+$ is positive. Then,
if $\widetilde{p}_+$ is negative (backward scattering), the post-bounce velocity $\widetilde P$ is smaller than the one obtained in the vacuum case,
whereas if it is positive (forward scattering), matter effects enlarge the value of $\widetilde P$.
This effect is also shown in Fig.~\ref{fig:post-bounce}, where both scenarios are depicted.
More specifically, from \eqref{kasnermap_stiffmatter} it can be seen that
\begin{align}
\label{backward_scatt}
	\widetilde p_+<0 \quad \iff \quad\overline\theta\in\left(
	\arccos\left[-\frac{4}{5}\left(1+\frac{\kappa}{3\overline P^2}\rho_{0{\rm stiff}}\right)^{1/2}\right],
	2\pi-\arccos\left[-\frac{4}{5}\left(1+\frac{\kappa}{3\overline P^2}\rho_{0{\rm stiff}}\right)^{1/2}\right]
	\right).
\end{align}
Thus, for this range of values, the scattering is backwards,
and the magnitude $\widetilde P$ is smaller than its vacuum counterpart,
ranging from $\widetilde P\in[P_{\min},P_{\rm mid}]$, where
\begin{align}\label{range_min_stiff_P}
	P_{\min}&:=\frac{\overline P}{3}
	\left[5-4\,\left(1+\frac{\kappa}{3\overline P^2}\rho_{0{\rm stiff}}\right)^{1/2}
	\right],
\\[5pt]
\label{range_mid_stiff_P}
P_{\rm mid}&:=\frac{3\overline P}{5}\left(
	1-\frac{16\kappa}{27\overline P^2}\rho_{0{\rm stiff}}
	\right)^{1/2}.
\end{align}
For sufficiently large ratios $\kappa\rho_{0{\rm stiff}}/\overline P^2$,
that is, for $\kappa\rho_{0\rm stiff}>27\overline P^2/16$,
this range becomes empty,
implying that the bounce always results in a forward scattering with $\widetilde p_+>0$. Forward scatterings occur for pre-bounce angles $\overline{\theta}$ not included in the interval \eqref{backward_scatt}, and
$\widetilde P$ ranges from $\widetilde P=P_{\rm mid}$ to $\widetilde P=\overline P$. In any case,
as for vacuum, the post-bounce norm is always smaller than the pre-bounce one.

Similarly to the vacuum scenario, the maximum change in the norm and angle of the velocity vector occurs for a head-on collision with $\overline{\theta} = \pi$. In this case, the post-bounce values are $\widetilde{P} = P_{\min}$ and $\widetilde{\theta} = 0$. In contrast, the minimum change occurs at the boundary angles $\overline{\theta} = \theta_{\rm stiff}$ and $\overline{\theta} = 2\pi - \theta_{\rm stiff}$, where the Kasner map \eqref{trans_P_stiff}--\eqref{trans_theta_stiff_2} yields the identity.

Furthermore, it is interesting to study at what point on the anisotropy plane $(\beta_+,\beta_-)$
the bounce occurs in each scenario.
By simply evaluating the general solutions \eqref{sol_beta_-_exact} and \eqref{sol_beta_+_vacuum_stiff} for the shape parameters at $\tau=0$, one obtains
\begin{align*}
	(\beta_+,\beta_-)|_{\rm bounce}=\left(\frac{C_1}{3}-\frac{1}{3}\ln(\rzero C_3),k_-\right).
\end{align*}
In this expression,
the sole difference between the vacuum and stiff-matter solutions is encoded in $C_3$, which for vacuum reads
as $C_3 = 2\overline{p}_+ - \overline{P}$, while for stiff matter is given by $C_3 = 2\overline{p}_+ - \overline{P} \left(1 + \frac{\kappa \rho_{0\rm stiff}}{3\overline{P}^2}\right)^{1/2}$.
Therefore, the matter density makes this constant smaller and, as a result, in the matter solution,
the bounce occurs at the same $\beta_-$ value but at greater value of $\beta_+$ as compared to the vacuum case;
that is, closer to the origin of the $(\beta_+,\beta_-)$ plane.
In particular this implies that the value of the potential at the bounce is lower for the matter case,
a feature that can also be observed in Fig.~\ref{fig:post-bounce}.

Moreover, returning to the full Bianchi IX scenario, as argued for the vacuum case, this bounce applies to the left wedge and for pre-bounce angles $\widetilde{\theta} \in (\theta_{\rm stiff}, 2\pi - \theta_{\rm stiff})$.
By performing a $2\pi/3$ clockwise and counter-clockwise rotation (due to the symmetry of the potential), we deduce the application range for the rest of the sectors: in the upper-right wedge for $\overline{\theta} \in (\theta_{\rm stiff} - 2\pi/3, 4\pi/3 - \theta_{\rm stiff})$, and in the lower-right wedge for $\overline{\theta} \in (\theta_{\rm stiff} + 2\pi/3, 8\pi/3 - \theta_{\rm stiff})$.
For angles outside these ranges, including the directions of the symmetry semiaxes $\overline{\theta} = 0, 2\pi/3, 4\pi/3$, a solution is asymptotically allowed, but no transition occurs, thus following an uninterrupted Bianchi I solution approaching the singularity. Then, considering the analogy previously mentioned for the vacuum case, it can be seen as if the channels through which the system escapes (the symmetry semiaxes in vacuum) are widened.

To finish, it is important to mention that in the full Bianchi IX dynamics, multiple bounces occur consecutively. Since the norm $P$ is reduced after each bounce, at some point the ratio  $\kappa\rho/\overline P^2$ grows large enough that the range of pre-bounce angles becomes empty, specifically when condition \eqref{max_rho_stiff} is saturated. Consequently, the sequence of bounces eventually ceases (in a finite time).
	Thus, for any given initial data, the system will eventually undergo a last bounce, and then follow the corresponding
	Bianchi I trajectory until reaching the singularity (see Refs.~\cite{Ringstrom:2000mk,Jantzen:2001me} for more details).

\begin{figure}
    \centering
    
    \begin{subfigure}[b]{0.3\textwidth}
        \centering
        \includegraphics[width=\linewidth]{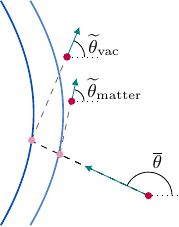}
    \end{subfigure}
\hspace{4cm}
    \begin{subfigure}[b]{0.35\textwidth}
        \centering
        \includegraphics[width=\textwidth]{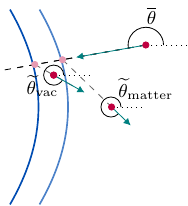}
    \end{subfigure}
\\
\vspace{0.5cm}
        \begin{subfigure}[b]{0.28\textwidth}
        \centering
        \includegraphics[width=\linewidth]{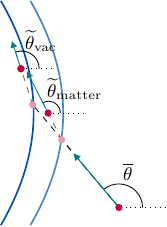}
    \end{subfigure}
    \hspace{4cm}
    \begin{subfigure}[b]{0.32\textwidth}
        \centering
        \includegraphics[width=\textwidth]{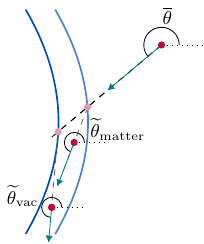}
    \end{subfigure}
    \caption{Comparison of a bounce against a potential wall in vacuum and with matter.
    The upper plots correspond to a backward scattering, where the horizontal component
    of the velocity vector changes sign. In these cases the norm of the post-bounce vector is smaller
    with matter than in the vacuum case. In contrast, the lower plots show a forward scattering,
    where the presence of matter increases the norm of the scattered velocity.
    In all the cases the deflection angle $\Delta\theta:=|\widetilde\theta-\overline\theta|$
    is smaller with matter, signaling an effective increase of the convex curvature of the wall.}
    \label{fig:post-bounce}
\end{figure}

\section{Kasner map for a general barotropic fluid}\label{sec.generalkasner}

Since, for $\omega\neq 1$, the contribution from the matter terms are
asymptotically  small, and the system tends to the vacuum dynamics, here we will linearize the equations of motion
around the vacuum solution. In this way, we will be able to solve the system
and explicitly provide the Kasner transition map between consecutive kinetic-dominated periods.

\subsection{Approximate solution}

Let us recapitulate: the system is described by four independent dynamical variables, $\rho$, $\beta_+$, $\beta_-$, and $x$ (one could instead work with $\alpha=(x+\beta_+)/2$, though $x$ turns out to be more convenient for the following computations), with dynamical equations
\begin{align}
\dot{\rho}&= -\frac{3}{2}(1+\omega)(x+\beta_+)\rho,
\label{eq_to_linearize1}\\\label{eq_to_linearize2}
\ddot{x}&=
\kappa (1-\omega)\rho e^{3(x+\beta_+)},
\\\label{eq_to_linearize3}
\ddot{\beta}_+&=\frac{4}{3\rzero^2}e^{2x-6\beta_+},
\\\label{eq_to_linearize4}
\ddot{\beta}_-&=0,
\end{align}
and subject to the constraint equation \eqref{constraint_II_x_y},
\begin{align}\label{constraint_to_lin}
\kappa\rho-3e^{-3(x+\beta_+)}\left[\frac{1}{4}\left(\dot{x}+\dot{\beta}_+\right)^2
-\dot{\beta}_+^2-p_-^2
\right]+\frac{1}{\rzero^2}e^{-x-9\beta_+}=0.
\end{align}
As commented above, we will assume that the evolution of these variables is given as their corresponding
vacuum solution (background) plus certain small perturbation, that is,
\begin{align}
\label{sol_gen_lin}
\begin{aligned}
x &=x^{(0)}+\delta x,\\
\rho &=\rho^{(0)}+\delta\rho,\\
\beta_+ &=\beta_+^{(0)}+\delta\beta_+,\\
\beta_-&=\beta_-^{(0)}+\delta\beta_-,
\end{aligned}
\end{align}
with the $(0)$ superindex indicating the vacuum solution, for which
$\rho^{(0)}=0$, and $x^{(0)}$, $\beta_+^{(0)}$, and $\beta_-^{(0)}$
are given by \eqref{sol_x_vacuum_stiff}, \eqref{sol_beta_+_vacuum_stiff}, and \eqref{sol_beta_-_exact}, respectively.
We now replace this decomposition in the equations of motion
\eqref{eq_to_linearize1}--\eqref{constraint_to_lin}, and linearize them
for the variables $\delta x$, $\delta\rho$, $\delta\beta_+$, and $\delta\beta_-$.
For $\rho$ and $\beta_-$ this is trivial since they already obey linear equations,
and thus their evolution is given by,
\begin{align}
\label{sol_delta_rho}
\delta\rho &=\rho_0 e^{-\frac{3}{2}(1+\omega)(x^{(0)}+\beta_+^{(0)})},
\\
\delta\beta_- &=\delta k_-+\delta p_-\,\tau.
\end{align}
The perturbation $\delta\beta_-$ simply implies a change of the background
constants of motion, and can thus be reabsorbed in the background
$\beta_-^{(0)}$ without loss of generality. This does not apply however to
$\delta\rho$, since its background $\rho^{(0)}$ is exactly vanishing.

Concerning the equations for $x$ \eqref{eq_to_linearize2} and $\beta_+$
\eqref{eq_to_linearize3}, it is immediate to obtain
\begin{align}
\label{eq_lin_x}
\ddot{\delta x}&=\kappa (1-\omega)e^{3(x^{(0)}+\beta_+^{(0)})}\delta \rho ,\\
\label{eq_lin_beta_+}
\ddot{\delta \beta_+}&=\frac{8}{3\rzero^2}e^{2(x^{(0)}-3\beta_+^{(0)})}(\delta x-3\delta\beta_+),
\end{align}
while from the constraint \eqref{constraint_to_lin} one gets the relation,
\begin{align}\label{constraint_linearized}
\begin{aligned}
0=&\frac{2}{\rzero^2}e^{2(x^{(0)}-3\beta_+^{(0)})}\left(
\delta x-3\delta \beta_+
\right)+
	\kappa e^{3(x^{(0)}+\beta_+^{(0)})}
\delta \rho
-\frac{3}{2}\left(
\dot x^{(0)}+\dot \beta_+ ^{(0)}
\right)
\dot{\delta x}
-\frac{3}{2}\left(
\dot x^{(0)}-3\dot \beta_+ ^{(0)}
\right)
\dot{\delta\beta_+}.
\end{aligned}
\end{align}
The solution of \eqref{eq_lin_x} for $\delta x$ can be directly obtained by integration, after
replacing \eqref{sol_x_vacuum_stiff}, \eqref{sol_beta_+_vacuum_stiff}, and \eqref{sol_delta_rho}, namely,
\begin{align}\label{sol_x}
\begin{aligned}
\delta x&=
-\frac{\kappa\rho_0 A}{32C_3^2}
e^{2(1-\omega)C_2\tau}
\left[2\cosh(2C_3\tau)\right]^{(5-\omega)/2}
+\frac{\kappa\rho_0 A(1-\omega)C_2}{32C_3^3}
\text{B}\!\left[
\frac{1}{2}(1+\tanh(2C_3\tau))
;b_1,
b_2
\right]
\\
&\quad
+\frac{\kappa\rho_0 A}{32C_3^2}(b_1-1)
\text{B}\!\left[
\frac{1}{2}(1+\tanh(2C_3\tau))
;b_1-1,
b_2+1
\right]
-\frac{\kappa\rho_0 A}{32C_3^2}(b_2-1)
\text{B}\!\left[
\frac{1}{2}(1+\tanh(2C_3\tau))
;b_1+1,
b_2-1
\right]
\\
&\quad
+\frac{\kappa\rho_0 A(1-\omega)}{16b_1^2 C_3^2}e^{(2C_2-C_3)(1-\omega)\tau}\,
{}_3\text{F}_2\left[\left\{b_1,b_1,\frac{\omega-1}{2}\right\};\{b_1+1,b_1+1\};
-e^{4C_3\tau}
\right],
\end{aligned}
\end{align}
where we have defined $b_1:=-\frac{1-\omega}{4C_3}
(C_3-2C_2)$, $b_2:=-\frac{1-\omega}{4C_3}
(C_3+2C_2)$, and $A:=(e^{4C_1}/(2C_3 \rzero))^{(1-\omega)/2}$. Moreover,
$\text{B}$ denotes the incomplete beta function, and ${}_3\text{F}_2$ the generalized hypergeometric function. Then, we can solve
\eqref{constraint_linearized} to obtain the solution for $\delta\beta_+$,
\begin{align}\label{sol_beta_+_1}
\begin{aligned}
\delta \beta_+&=
\frac{\kappa\rho_0 A}{32C_3^2}
e^{2(1-\omega)C_2\tau}
\left[2\cosh(2C_3\tau)\right]^{(5-\omega)/2}
-\frac{\kappa\rho_0 A(1+3\omega)}{96C_3^2}\tanh(2C_3\tau)
\text{B}\!\left[
\frac{1}{2}(1+\tanh(2C_3\tau))
;b_1-1,
b_2-1
\right]
\\
&\quad
-\frac{\kappa\rho_0 A(1-\omega)}{96C_3^3}\left[7C_2+8C_3\tanh(2C_3\tau)\right]
\text{B}\!\left[
\frac{1}{2}(1+\tanh(2C_3\tau))
;b_1,
b_2
\right]
\\
&\quad
+\frac{\kappa\rho_0 A}{96C_3^3}\left[
(b_1-1)C_3+2(1-\omega)
(C_2-2C_3)
\tanh(2C_3\tau)\right]	
\text{B}\!\left[
\frac{1}{2}(1+\tanh(2C_3\tau))
;b_1-1,
b_2+1
\right]
\\
&\quad
-\frac{\kappa\rho_0 A}{96C_3^3}\left[
(b_2-1)C_3+2(1-\omega)
(2C_3+C_2)
\tanh(2C_3\tau)\right]	
\text{B}\!\left[
\frac{1}{2}(1+\tanh(2C_3\tau))
;b_1+1,
b_2-1
\right]
\\
&\quad
+\frac{\kappa\rho_0 A(1-\omega)}{48b_1^2C_3^3}
e^{(2C_2-C_3)(1-\omega)\tau}	
\left[C_3+2C_2
\tanh(2C_3\tau)\right]\,
{}_3\text{F}_2\left[\left\{b_1,b_1,\frac{\omega-1}{2}\right\};\{b_1+1,b_1+1\};
-e^{4C_3\tau}
\right].
\end{aligned}
\end{align}
Finally, it is immediate
to write the perturbation of $\alpha$ in terms of the above:
\begin{equation}\label{sol_alpha_1}
\delta\alpha=\frac{1}{2}(\delta x+\delta\beta_+).
\end{equation}
We note that the constants $C_1$, $C_2$, and $C_3$ that appear in these expressions
correspond to the background solutions \eqref{sol_x_vacuum_stiff} and \eqref{sol_beta_+_vacuum_stiff}; in particular,
the case $C_3=0$ is not included since the background solution constrains
$C_3$ to be positive (see Eq.~\eqref{solq}). Additionally,
the integration constants that come from solving the linearized system \eqref{eq_lin_x}--\eqref{eq_lin_beta_+}
have been reabsorbed in the background quantities without loss of generality, as we did above for $\delta\beta_-$.

As it can be seen, the corrections $\delta x$ and $\delta\beta_+$
(and therefore also $\delta\alpha$) to the vacuum solutions
are scaled by the constant $\kappa\rho_0 A$. It is possible
to see that this constant is related to the value of $\alpha$
and $\delta\rho$ at the transition $\tau=0$.\footnote{
More specifically, 
by direct evaluation of $\rho e^{6\alpha}$ at the bounce in $\tau=0$, making use of the equations \eqref{sol_delta_rho} and \eqref{sol_alpha_vacuum_stiff}, to first order in the linearization, we have that
$\rho e^{6\alpha}|_{\tau=0}=\rho_0\left(e^{4C_1}/
{(C_3\rzero)}
\right)^{\!\frac{1-\omega}{2}}$,
and thus $\kappa\rho_0A$ is proportional to $\rho e^{6\alpha}|_{\tau=0}$.
}
More precisely, the smaller the value of $\alpha(0)$ and $\delta\rho(0)$,
the smaller value $\kappa\rho_0 A$ will take.
Therefore, as one would expect, the matter corrections diminish as the singularity $\alpha\to-\infty$ is approached
or as the matter density on the transition tends to zero.

\subsection{Kasner map}

Once that we have obtained the evolution of the variables,
we are now in a position to compute the Kasner map
that relates two subsequent (pre- and post-bounce) Bianchi I periods.
Following the prescription we applied for vacuum, we choose the singularity to be located at $\tau\to-\infty$,
and thus the branch $\tau \to +\infty$ corresponds to the pre-bounce, while $\tau \to -\infty$ corresponds to the post-bounce, kinetic-dominated epochs.

Now we need to obtain an approximate version of the Bianchi I dynamics to fit with the obtained solutions
at $\tau \to\pm\infty$. For such a purpose, one can set up a similar linearization as \eqref{sol_gen_lin}
for a Bianchi I vacuum background, which, in particular, implies $\rzero\to\infty$. The matter density
$\delta\rho$ has the same form as \eqref{sol_delta_rho}, the shape parameters are linear in $\tau$,
and one simply needs to solve \eqref{eq_lin_x} to obtain the corresponding evolution for $\alpha$.
All in all, this quite straightforward computation leads to the form
\begin{align}\label{evol_Bianchi_I_approx}
\begin{aligned}
\alpha&=P \tau+c_\alpha+\frac{\kappa\rho_0}{18P^2(1-\omega)}e^{3(1-\omega)(P\tau+c_\alpha)},
\\
\beta_+&=p_+ \tau+ k_+,
\\
\beta_-&=p_- \tau+k_-,
\end{aligned}
\end{align}
where $\{p_+,p_-,k_+,k_-,c_\alpha\}$ are again the five Bianchi I parameters characterizing
the kinetic-dominated periods.

Following a similar procedure as applied in Subsec.~\ref{sec.vacuum.kasner} for the vacuum case,
the parameters $\{p_+,p_-,k_+,k_-,c_\alpha\}$ that characterize the dynamics in the (pre- and post-bounce) branch can be obtained in terms of the constants $C_i$ of the general solution \eqref{sol_x}--\eqref{sol_alpha_1}.
Then, inverting such relations, it is possible to obtain
the Kasner map that relates the pre-bounce state  $\{\overline p_+,\overline p_-,\overline k_+,\overline k_-,\overline c_\alpha\}$,
denoted with an overline, with the post-bounce state $\{\widetilde p_+,\widetilde p_-,\widetilde k_+,\widetilde k_-,\widetilde c_\alpha\}$, denoted by a tilde:
\begin{align}\label{transition_law_lin}
\begin{aligned}
\widetilde p_+&=\frac{1}{3}(4\overline P-5\overline p_+)
+\frac{2\varepsilon \overline P}
{9}
\frac{\Gamma\left(
	1+\mu
	\right)
	\Gamma\left(
	1+\nu
	\right)}{\Gamma\left(\frac{1+\omega}{2}\right)},
\\
\widetilde k_+&=
\overline k_+
+
\frac{\varepsilon \overline P^2}
{48(\overline P-2\overline p_+)^3}
\frac{\Gamma(\mu)\Gamma(\nu)}{\Gamma\left(
	\frac{\omega-1}{2}
	\right)}
\left[
4(\overline P-2\overline p_+)
-3(1-\omega)\overline P
\left(
2+
\frac{\Gamma'(\nu)}{\Gamma(\nu)}
-\frac{\Gamma'(\mu)}{\Gamma(\mu)}
\right)
\right],
\\
\widetilde c_\alpha&=
\overline c_\alpha
+
\frac{\varepsilon \overline P^2}
{48(\overline P-2\overline p_+)^3}
\frac{\Gamma(\mu)\Gamma(\nu)}{\Gamma\left(
	\frac{\omega-1}{2}
	\right)}
\left[
2(\overline P-2\overline p_+)
-3(1-\omega)
\left(
\overline P+\overline p_+
\frac{\Gamma'(\nu)}{\Gamma(\nu)}
-\overline p_+\frac{\Gamma'(\mu)}{\Gamma(\mu)}
\right)
\right],\\
\widetilde p_-&=\overline p_-,\\
\widetilde k_-&=\overline k_-,
\end{aligned}
\end{align}
where we have defined $\mu:=\frac{3(1-\omega)\overline P}{4(\overline P-2\overline p_+)}$,
$\nu:=-(1-\omega)\frac{5\overline P-4\overline p_+}{4(\overline P-2\overline p_+)}$,
$\varepsilon:=\kappa \rho_0e^{3(1-\omega)\overline c_\alpha}/\overline P^2$,
and $\overline P=(\overline p_+^2+\overline p_-^2)^{1/2}$. Moreover, $\Gamma$
denotes the gamma function, and here the prime stands for a derivative with respect to the argument.

This is the main result of the present paper and explicitly
shows how the leading matter terms affect the relation between the parameters
corresponding to consecutive kinetic-dominated periods.
This Kasner map completes and generalizes the transition law presented in Ref.~\cite{Ali:2017qwa}, where the particular case of a dust field ($\omega=0$) was considered
and the transition law was provided only for a subset of the dynamical variables.
Note that, as commented above, in vacuum --- with the transition law given by \eqref{transition_stiff_vacuum} --- 
the constants $k_+$, $k_-$, and $c_\alpha$ are irrelevant, as they are conserved through the bounce and can be absorbed in a global (time-independent)
change of coordinates. However, under the presence of matter,
this is no longer possible; the value of these parameters, which describe
the pre- and post-bounce Bianchi I epochs,
change at the bounce, and thus it is not possible to perform
a global change of coordinates to absorb them.
Therefore, in order to have the complete information of the dynamics,
it is necessary to provide their corresponding transition law.

Let us now analyze the validity of the map \eqref{transition_law_lin}.
The background solution \eqref{sol_alpha_vacuum_stiff}--\eqref{sol_beta_+_vacuum_stiff} already excludes the case $\overline P=2\overline p_+$, due to the requirement that $C_3=2\overline p_+-\overline P$ be strictly positive.
In addition, from the definition of $\varepsilon$ it is clear that $\overline P$ cannot be zero; however, this does not exclude any case of interest,
	as it corresponds to the isotropic case
	where there are no bounces.
However, the gamma functions diverge if their argument is a nonpositive integer,
and this fact will determine the limits of validity for the obtained solution. In order to analyze these divergences, it is convenient to parametrize the pre-bounce
velocity in terms of its polar components $(\overline P, \overline\theta)$, as introduced in the
previous section, which leads to the following form of the arguments of the gamma functions,
\begin{align}
\label{def_mu_theta}
\mu &= \frac{3(1-\omega)}{4(1+2 \cos\overline\theta)},\\
\label{def_nu_theta}
\nu &= -\frac{(1-\omega)}{4}\left(\frac{5+4\cos\overline\theta}{1+2\cos\overline\theta}\right).
\end{align}
Since $1-\omega>0$ and the background solution restricts $\overline\theta$ to angles $\overline\theta\in(2\pi/3, 4\pi/3)$, which implies $\cos\overline\theta<-1/2$, it is clear
that $\mu$ is negative definite, while $\nu$ is positive definite. Therefore, when $\mu$ is a negative integer,
\begin{equation}\label{eqm}
\mu=-m,
\end{equation}
with $m\in\mathbb{N}-\{0\}$, the functions $\Gamma(\mu+1)$, $\Gamma(\mu)$, and $\Gamma'(\mu)$ will be divergent.
For each integer $m$, from \eqref{eqm} the corresponding angle $\overline\theta_m$
can be obtained,
\begin{equation}\label{cond_divergence}
\cos\overline\theta_m=-\frac{1}{2}-\frac{3}{8m}(1-\omega),
\end{equation}
that leads to a pole of the gamma function.
Since $m$ can be any positive integer, this result implies that there is an infinite number of pre-bounce angles $\overline\theta$
for which the Kasner map diverges. Specifically, starting from $\cos\overline{\theta} = -\frac{1}{2}$
(that corresponds to $m\to+\infty$) and gradually decreasing its value, there are infinitely many values of $\cos\overline{\theta}$ where condition \eqref{cond_divergence} is satisfied, until reaching $\cos\overline\theta=-\frac{1}{2}-\frac{3}{8}(1-\omega)$,
where the last singularity (corresponding to $m=1$) occurs. Therefore, in the range
\begin{equation}
-1\leq \cos\overline\theta<-\frac{1}{2}-\frac{3}{8}(1-\omega),
\end{equation}
or, equivalently,
\begin{equation}\label{range_validity_theta_lin}
\arccos\left(-\frac{1}{2}-\frac{3}{8}(1-\omega)\right)\leq \overline\theta\leq2\pi-\arccos\left(-\frac{1}{2}-\frac{3}{8}(1-\omega)\right),
\end{equation}
there are no poles of the gamma functions.
Thus, we conclude that this is the range of pre-bounce angles for which the approximation is valid.
As can be observed, this interval is centered around $\overline\theta=\pi$, and
its length depends on $\omega$: it widens as $\omega$ approaches one, while it is empty for $\omega < -1/3$. Thus, the applicability of the Kasner map \eqref{transition_law_lin} is also restricted to $-1/3\lesssim\omega< 1$. Therefore, taking this into account, one can expect the Kasner map to yield sensible results for angles within a relatively small
interval around $\overline{\theta}=\pi$, with the length of the interval being wider for matter fields with a higher barotropic index. 

In particular, it is interesting to note that, although the stiff-matter case $(\omega = 1)$ was not included in the present perturbative analysis (as the solutions \eqref{sol_beta_+_1} and \eqref{sol_alpha_1} do not include it), expanding the exact Kasner map \eqref{transition_stiff_vacuum} of $p_+$ for stiff matter at linear order in $\kappa \rho_{0{\rm stiff}}$ exactly reproduces the result of \eqref{transition_law_lin}, by simply identifying $\varepsilon$ with $\kappa \rho_{0{\rm stiff}} / \overline{P}^2$. For the transition laws of $k_+$ and $c_\alpha$ in \eqref{transition_law_lin}, there is a divergence as $\omega\to 1$,
as one can check by expanding the arguments of the gamma functions around $\omega=1$.
However, this is not a divergence that limits the applicability of this map.
Note that one could simply rescale the constant $c_\alpha$ in \eqref{evol_Bianchi_I_approx}
in a suitable manner so that the correction terms are also rescaled, and thus no such divergence
appears. (Concerning $k_+$, we remind that $k_+$ and $c_\alpha$ are related, as in the background case by \eqref{rel_k_+_alpha}, and thus
any issues with $\widetilde k_+$ would in principle be resolved by avoiding the singularities of $\widetilde c_\alpha$.) Also, as we will explain below, larger values of $c_\alpha$ correspond to the bounce taking place closer to the singularity.

Let us now comment some general features of the Kasner map. All the terms that appear due to
the presence of matter fields are proportional to the factor $\varepsilon=\kappa\rho_0 e^{3(1-\omega)\overline c_\alpha}/\overline P^2$.
Therefore, by construction, the limit $\varepsilon\to 0$ yields the well-known vacuum Kasner map
\eqref{transition_stiff_vacuum}. This limit corresponds either to an exactly vanishing
contribution of matter fields, i.e., $\rho_0=0$, or to $\overline c_{\alpha}\to-\infty$, which implies
$e^{3\overline c_{\alpha}(1-\omega)}\to 0$. The constant $\overline c_{\alpha}$ is related to the value of $\alpha$ at the bounce,\footnote{Specifically, since the bounce occurs at $\tau=0$, according to the background solution \eqref{sol_alpha_vacuum_stiff}, we have that $e^{6\alpha(0)}=e^{4C_1}/(\rzero C_3)$ up to small matter corrections. Then,
by identification with the pre-bounce Bianchi I dynamics \eqref{Kasner_param_ini},
$\overline c_\alpha\approx\alpha(0)-\frac{1}{6}\ln2$.} and, in particular, 
$\overline c_\alpha\to -\infty$ can be understood
as the limit when the bounce happens very close to the singularity.

It turns out that
the leading matter terms for a generic barotropic fluid have similar qualitative effects
as those described in Subsec.~\ref{sec.kasner.stiff} for a stiff-matter content. First of all, by evaluating the solution of $\beta_+$ \eqref{sol_beta_+_1} at the bounce, it can be seen that its value is larger than the corresponding to the vacuum case. Then, the bounce occurs
at larger values of $\beta_+$, and thus closer to the origin in the $(\beta_+,\beta_-)$ plane, and for smaller values of the potential. Moreover, the post-bounce value of the component $p_+$ is greater than in the vacuum case because, within the range of validity of the map, the $\Gamma$ functions are strictly positive, ensuring that the matter corrections in \eqref{transition_law_lin} are always positive. Consequently, since the component $p_-$ is conserved, for each pre-bounce angle $\overline{\theta}$, the deflection angle $\Delta \theta := |\widetilde{\theta} - \overline{\theta}|$ is always smaller than in vacuum, decreasing further with the fraction $\varepsilon$, which measures the strength of the matter terms. 
This can clearly be observed in Fig.~\ref{fig:delta_theta}, where the post-bounce angle $\widetilde\theta$
is plotted in terms of the pre-bounce one $\overline\theta$ for different values of the parameter $\varepsilon$.
As can be seen in this figure, for a head-on collision with $\overline{\theta}=\pi$, the matter
effects exactly vanish, resulting in a post-bounce angle $\widetilde{\theta}=0$, regardless of the matter content.
This outcome arises from the conservation of the momenta $p_-$: if it is zero in the pre-bounce state ($\overline{\theta}=\pi$), it must remain zero after the bounce as well (then $\widetilde\theta=0$). However, as the pre-bounce angle departs from this $\overline{\theta}=\pi$ value,
the matter effects increase. Intuitively, this can be understood as
matter effectively increasing the convex curvature of the potential wall,
as can be seen in Fig.~\ref{fig:post-bounce}, which also applies to this more general matter content}. For the norm of the velocity $P$, the same argument applies as with stiff matter: when the scattering is forward ($\widetilde{p}_+ > 0$), the norm increases as
compared to vacuum, whereas for a backward scattering ($\widetilde{p}_- < 0$), the norm decreases.

\begin{figure}
	\centering
	\includegraphics[width=0.8\linewidth]{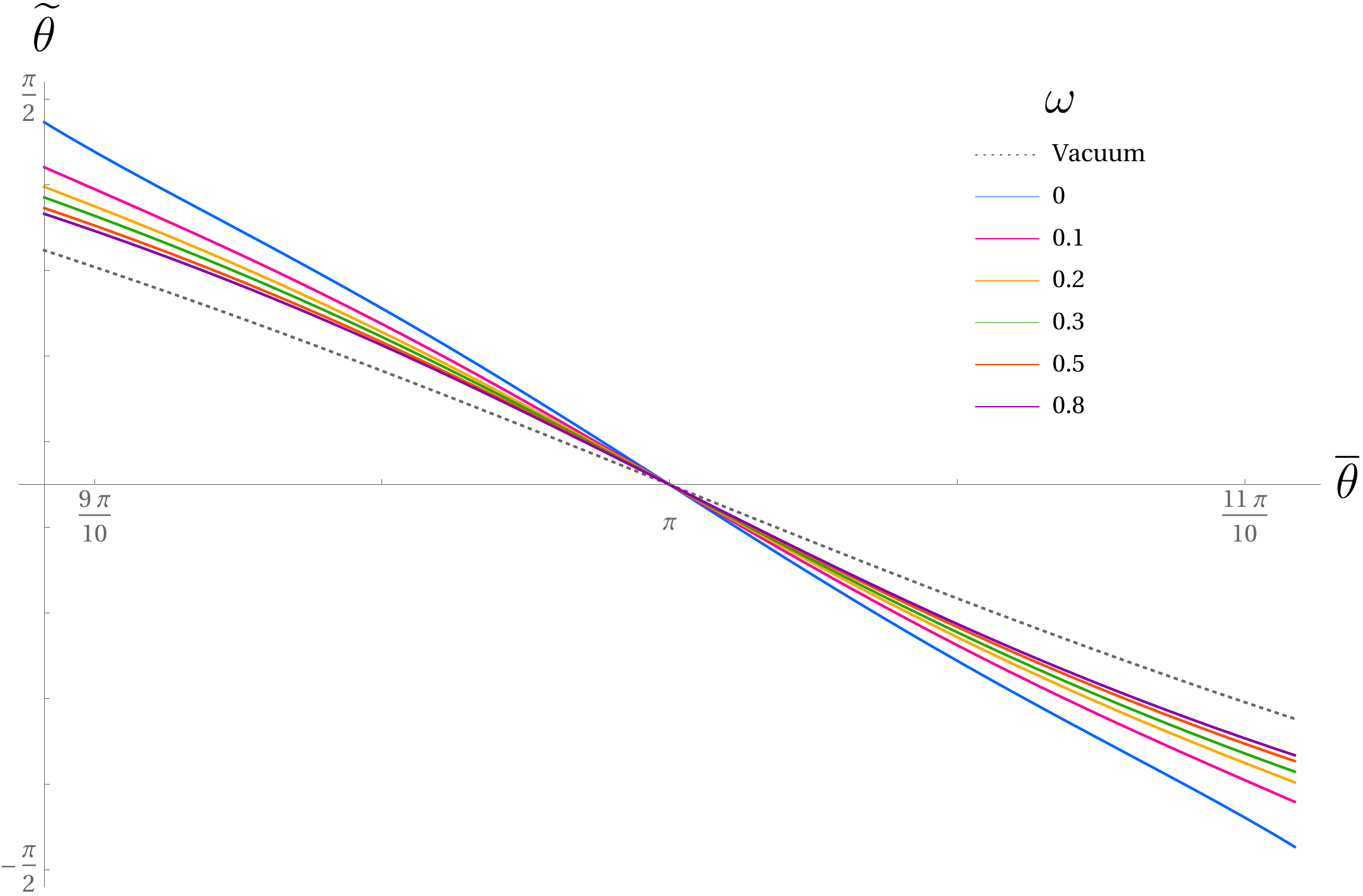}
	\caption{In this plot the post-bounce angle $\widetilde\theta$ is depicted in terms of the pre-bounce angle $\overline\theta$, obtained from the transition law \eqref{transition_law_lin} for a fixed value of $\varepsilon=0.2$ and varying values of the barotropic index $\omega$. For a more complete analysis on the influence of matter, the transition law corresponding to vacuum has also been plotted.}
	\label{fig:thetamatter}
\end{figure}

Furthermore, it is also interesting to study the influence of the barotropic index in the transition law \eqref{transition_law_lin}. Given that the analytic expressions involve several gamma functions, and are thus
difficult to interpret, in Fig.~\ref{fig:thetamatter} we depict the post-bounce angle $\widetilde{\theta}$ as a function of the pre-bounce angle $\overline{\theta}$, for a fixed value of the parameter $\varepsilon$ and varying $\omega$. To make the analysis more complete, we have also plotted the transition law corresponding to the vacuum case. To avoid singularities for the smaller barotropic indices, the range of pre-bounce angles $\widetilde{\theta}$ must be significantly restricted, and that is why it is reduced to a small range around $\pi$. As mentioned earlier, the validity range \eqref{range_validity_theta_lin} of the Kasner map \eqref{transition_law_lin} diminishes as $\omega$ decreases, leading us to exclude very small barotropic indices. From this result, the main observation is that the deflection angle $\Delta\theta:=|\widetilde\theta-\overline\theta|$ decreases as $\omega$ decreases. However, regardless of the value of $\omega$ and excluding a head-on collision, $\Delta\theta$ is always smaller than in vacuum. Building upon our earlier discussion, we conclude that a smaller barotropic index $\omega$, or a larger $\rho_0$, results in a greater enhancement of the convex curvature of the potential wall.

\begin{figure}
	\centering
	\includegraphics[width=0.85\linewidth]{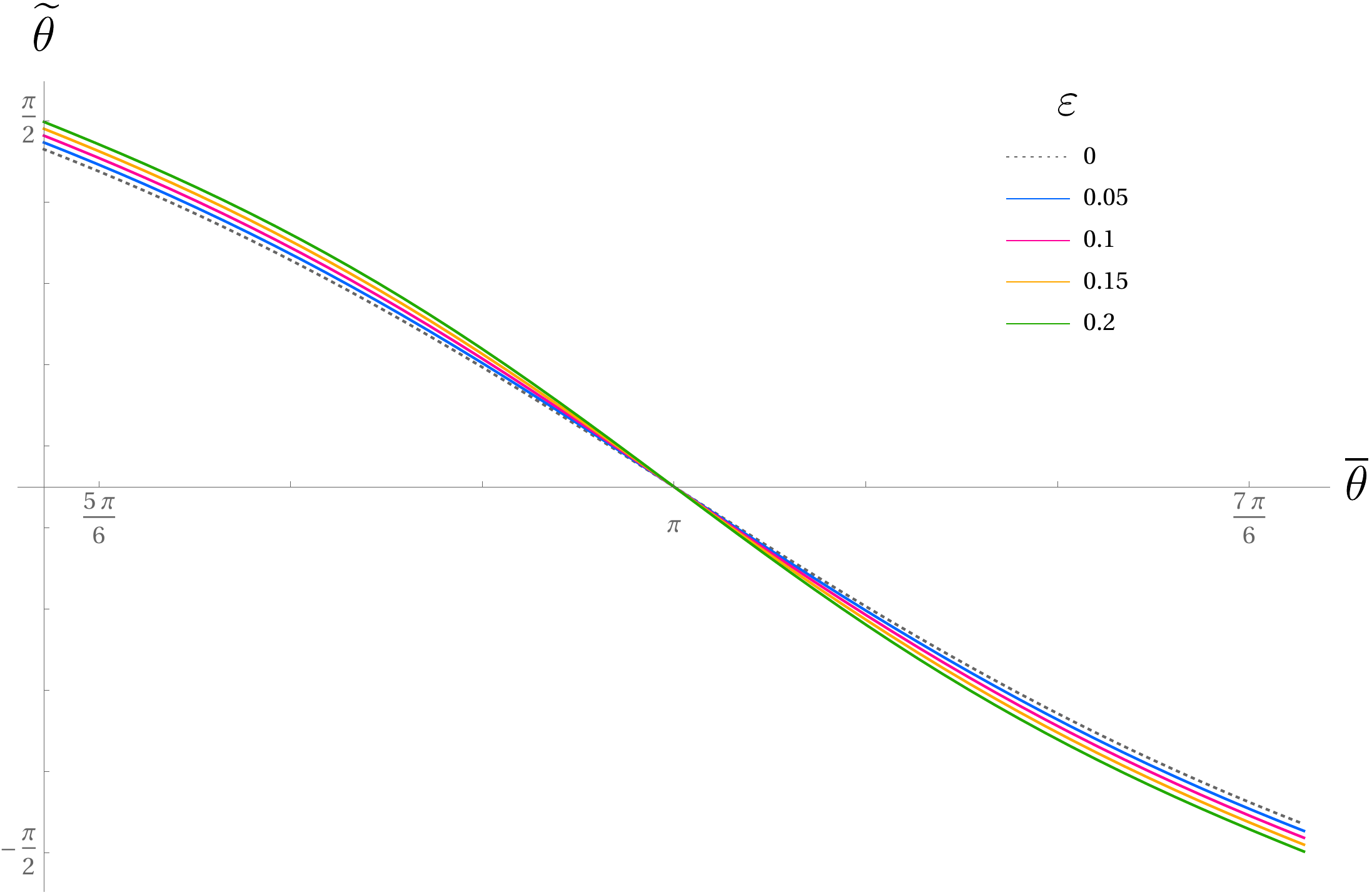}
	\caption{
		In this plot the post-bounce angle $\widetilde\theta$ is depicted in terms of the pre-bounce angle $\overline\theta$, obtained from the transition law \eqref{transition_law_lin} and with different values of the parameter $\varepsilon:=\kappa\rho_0e^{3(1-\omega)\overline c_\alpha}/\overline P^2$, which measures the strength of the matter terms.}
	\label{fig:delta_theta}
\end{figure}

\section{Asymptotics of the Kretschmann curvature invariant}
\label{sec:curvatureBIX}

Finally, let us check the asymptotic behavior
of the Kretschmann curvature scalar toward the singularity.
This scalar is defined as
	\begin{equation}
	\label{def_Kretschmann}
	K=R_{\mu\nu\lambda\rho}R^{\mu\nu\lambda\rho},
	\end{equation}
	with $R_{\mu\nu\lambda\rho}$ being the Riemann tensor, and it is usually
	used to characterize the blow-up of the curvature at the singularities.

Since in this regime
the system spends most of the time in kinetic-dominated periods,
we will compute its behavior during one such period,
though it will show rapid changes of behavior during the interaction
of the system with the potential walls.
Therefore, we simply replace the kinetic-dominated evolution with matter
for the variables, given by Eq.~\eqref{evol_Bianchi_I_approx}, in the above definition, and then take the limit toward the singularity $\tau\to -\infty$.

On the one hand, for the isotropic case $P=0$, with $\rho_0>0$
	and $\omega\in[-1,1)$, the Kretschmann scalar scales as a power-law in $\tau$, more specifically,
\begin{align}\label{kretschmann_iso_bianchi_IX}
K\propto\begin{cases}
|\tau|^{8/3(1-\omega)},& {\rm for}\,\,\,\, \omega\in\left[-1,-\frac{1}{3}\right],
\\[10pt]
|\tau|^{4(1+\omega)/(1-\omega)},& {\rm for}\,\,\,\, \omega\in\left[-\frac{1}{3},1\right).
\end{cases}
\end{align}
On the other hand, for the generic anisotropic case $P\neq 0$, with
$\rho_0\geq0$ and $\omega\in[-1,1]$, the scaling of the Kretschmann scalar reads,
\begin{align}\label{kretschmann_bianchi_ix}
K\propto\begin{cases}
e^{-4P\tau(1+2\cos\theta+2\sqrt{3}\sin\theta)},& {\rm for}\,\,\,\,  \theta\in\left[0,\frac{2\pi}{3}\right),
\\[10pt]
e^{-4P\tau(1-4\cos\theta)},& {\rm for}\,\,\,\,  \theta\in\left[\frac{2\pi}{3},\frac{4\pi}{3}\right),
\\[10pt]
e^{-4P\tau(1+2\cos\theta-2\sqrt{3}\sin\theta)},& {\rm for}\,\,\,\,  \theta\in\left[\frac{4\pi}{3},2\pi\right),
\end{cases}
\end{align}
which strongly depends on the angle $\theta$.
In particular, it is easy to see that all the terms in round parentheses
in the exponents of the last expression take values in the range $[3,5]$.
This leads to a divergence of the leading term of the
Kretschmann scalar between $e^{-12 P\tau}$ and $e^{-20 P\tau}$. More precisely, the slowest divergence rate $e^{-12 P\tau}$
takes place for the angles $\theta=0$, $2\pi/3$, and $4\pi/3$,
that correspond to the trajectories
that are following a kinetic-dominated period until the singularity
and will not interact with the potential walls, as commented in Sec.~\ref{sec.vacuum}. The further $\theta$ is
from the commented angles, the faster $K$ will diverge toward the singularity.
In fact, the maximum divergence rate of $e^{-20 P\tau}$
corresponds to the angles $\theta=\pi/3$, $\pi$, and $5\pi/3$, which define
trajectories that eventually
will undergo head-on collisions against the potential wall.
In any case, for the Bianchi IX geometry, the leading term of the Kretschmann scalar
diverges at least as fast as for the exact Bianchi I geometry,
where it scales as $e^{-12P\tau}$.

However, it is worth noting that the proportionality constant that, for the sake
of clarity, we have refrained from writing in \eqref{kretschmann_bianchi_ix},
can be vanishing for certain trajectories. For such solutions the scaling is slower,
but there are always subdominant terms
that, even in the vacuum case, make the Kretschmann scalar to diverge.

\section{Conclusions}\label{sec:conclusions}

We have presented a detailed asymptotic analysis of the Bianchi IX
cosmology coupled to matter. It is well known that, excluding stiff
matter (which is equivalent to a massless scalar field), the effects of the matter fields are generically negligible as
the singularity is approached, and thus the system
follows the same qualitative dynamics as in vacuum. However,
this is true only in the exact asymptotic limit
(in our gauge the initial singularity is located at $\tau\to-\infty$),
and, for finite values of time, the matter does indeed affect
the evolution of the universe. The main focus of our study
has been on specifically determining such effects.

The Bianchi IX dynamics can
be described as a succession of kinetic-dominated evolution periods, when
the system follows a Bianchi I dynamics, interrupted by quick
interactions with the potential walls. Our main goal in this
context has been to obtain the Kasner map, which relates the parameters
of two consecutive kinetic-dominated periods, taking into account
the leading effects of the matter fields.
Therefore, two main assumptions have been
implemented. First, the Bianchi IX potential has been approximated
as a pure exponential.
Under such assumption, for vacuum it is possible to obtain exact solutions of the
dynamics and, comparing the asymptotic (pre- and post-bounce) states,
obtain the Kasner map \eqref{transition_stiff_vacuum}.
Second, we have assumed that, even if there may be several
matter species in the model, the matter contribution
in the region close to the singularity can be described as a perfect
fluid with a linear equation of state.
Under the presence of such generic matter, the solution of
the equations of motion is not at hand.

Hence, we have performed
a linearization of the equations expanding the dynamical variables
around their form in the vacuum case. Since, toward the singularity,
the system dynamically tends to the vacuum solution,
such expansion can be understood as an expansion around
small volumes or, equivalently, around small matter densities.
In this way, we have been able to analytically solve the equations
of motion and, comparing the asymptotic states, derive the Kasner map
\eqref{transition_law_lin} that includes the leading effects of the matter fields.
This Kasner map can be used to approximate the complex Bianchi IX
dynamics in a discrete manner, and thus study
the chaotic nature of the system under the influence of matter fields, following, for instance, the work of Ref.~\cite{Cornish:1996hx} for vacuum. 

The specific quantitative form of the matter effects depends strongly on the barotropic index and
the state parameters that describe the velocity of the system in the plane of anisotropies
(see Figs.~\ref{fig:delta_theta}--\ref{fig:thetamatter}),
being minimum for a head-on collision, when the velocity is parallel to the $\beta_+$-axis
in the plane of anisotropies.
However, we have been able to show certain generic effects that can be clearly seen
in Fig.~\ref{fig:post-bounce}. In particular, generically
for forward (backward)
scatterings the norm of the scattered velocity is increased (reduced) by matter effects.
In addition, the presence of matter produces
the bounce at a lower value of the potential, and leads to a scattered velocity
with a lower deflection angle than in the vacuum case. That is, effectively matter
increases the convex curvature of the potential walls, which may have important consequences
for the chaotic nature of the system, as convex, and thus defocusing walls, are a necessary
condition for the presence of chaos in this kind of bouncing systems.

Finally, we have analyzed the scaling of the Kretschmann curvature invariant
toward the singularity, and explicitly derive its dependence
with the polar angle of the velocity in the plane
of anisotropies. All in all,
our analysis points out that, even if matter is completely negligible
in the exact limit of the singularity, its effects can be of high relevance
in the evolution of the system toward such limit.

\section*{Acknowledgments}
This work is supported by
the Basque Government Grant \mbox{IT1628-22}, and
by the Grant PID2021-123226NB-I00 (funded by MCIN/AEI/10.13039/501100011033 and by “ERDF A way of making Europe”).
SFU is funded by an FPU fellowship of the Spanish Ministry of Universities.

\bibliographystyle{bib-style}
\bibliography{references}

\providecommand{\href}[2]{#2}\begingroup\raggedright\begin{thebibliography}{10}

\bibitem{Belinsky:1970ew}
V.~A. Belinsky, I.~M. Khalatnikov, and E.~M. Lifshitz, ``{Oscillatory approach
  to a singular point in the relativistic cosmology},'' Adv. Phys. {\bf 19}
  (1970) 525.

\bibitem{Berger:2002st}
B.~K. Berger, ``{Numerical approaches to space-time singularities},'' Living
  Rev. Rel. {\bf 5} (2002) 1, \href{http://arXiv.org/abs/gr-qc/0201056}{{\tt
  arXiv:gr-qc/0201056}}.

\bibitem{Berger:1993ff}
B.~K. Berger and V.~Moncrief, ``{Numerical investigations of cosmological
  singularities},'' Phys. Rev. D {\bf 48} (1993) 4676,
  \href{http://arXiv.org/abs/gr-qc/9307032}{{\tt arXiv:gr-qc/9307032}}.

\bibitem{Berger:1998vxa}
B.~K. Berger, D.~Garfinkle, J.~Isenberg, V.~Moncrief, and M.~Weaver, ``{The
  Singularity in generic gravitational collapse is space - like, local, and
  oscillatory},'' Mod. Phys. Lett. A {\bf 13} (1998) 1565,
  \href{http://arXiv.org/abs/gr-qc/9805063}{{\tt arXiv:gr-qc/9805063}}.

\bibitem{Berger:1998wr}
B.~K. Berger and V.~Moncrief, ``{Numerical evidence that the singularity in
  polarized U(1) symmetric cosmologies on T**3 x R is velocity dominated},''
  Phys. Rev. D {\bf 57} (1998) 7235,
  \href{http://arXiv.org/abs/gr-qc/9801078}{{\tt arXiv:gr-qc/9801078}}.

\bibitem{Berger:2014tev}
B.~K. Berger, {\em {Singularities in cosmological spacetimes}}, p.~437.
\newblock 2014.

\bibitem{Garfinkle:2003bb}
D.~Garfinkle, ``{Numerical simulations of generic singularities},'' Phys. Rev.
  Lett. {\bf 93} (2004) 161101, \href{http://arXiv.org/abs/gr-qc/0312117}{{\tt
  arXiv:gr-qc/0312117}}.

\bibitem{Garfinkle:2004ww}
D.~Garfinkle, ``{The nature of gravitational singularities},'' Int. J. Mod.
  Phys. D {\bf 13} (2004) 2261, \href{http://arXiv.org/abs/gr-qc/0408019}{{\tt
  arXiv:gr-qc/0408019}}.

\bibitem{Heinzle:2012um}
J.~M. Heinzle, C.~Uggla, and W.~C. Lim, ``{Spike oscillations},'' Phys. Rev. D
  {\bf 86} (2012) 104049, \href{http://arXiv.org/abs/1206.0932}{{\tt
  arXiv:1206.0932}}.

\bibitem{Misner:1969hg}
C.~W. Misner, ``{Mixmaster universe},'' Phys. Rev. Lett. {\bf 22} (1969) 1071.

\bibitem{Misner:1969ae}
C.~W. Misner, ``{Quantum cosmology. 1.},'' Phys. Rev. {\bf 186} (1969) 1319.

\bibitem{Barrow:1981prl}
J.~D. Barrow, ``{Chaos in the Einstein equations},'' Phys. Rev. Lett. {\bf 46}
  (1981) 963.

\bibitem{Barrow:1981sx}
J.~D. Barrow, ``{Chaotic behavior in general relativity},'' Phys. Rept. {\bf
  85} (1982) 1.

\bibitem{Chernoff:1983zz}
D.~F. Chernoff and J.~D. Barrow, ``{Chaos in the Mixmaster universe},'' Phys.
  Rev. Lett. {\bf 50} (1983) 134.

\bibitem{Hobill:1993osv}
D.~Hobill, A.~Burd, and A.~Coley, eds., {\em {Deterministic chaos in general
  relativity}: {proceedings, NATO advanced research workshop, Kananaskis,
  Canada, 25-30 Jul, 1993}}.
\newblock 1993.

\bibitem{Motter:2003jm}
A.~E. Motter, ``{Relativistic chaos is coordinate invariant},'' Phys. Rev.
  Lett. {\bf 91} (2003) 231101, \href{http://arXiv.org/abs/gr-qc/0305020}{{\tt
  arXiv:gr-qc/0305020}}.

\bibitem{Imponente:2001fy}
G.~Imponente and G.~Montani, ``{On the covariance of the Mixmaster
  chaoticity},'' Phys. Rev. D {\bf 63} (2001) 103501,
  \href{http://arXiv.org/abs/astro-ph/0102067}{{\tt arXiv:astro-ph/0102067}}.

\bibitem{Cornish:1996yg}
N.~J. Cornish and J.~J. Levin, ``{The Mixmaster universe is chaotic},'' Phys.
  Rev. Lett. {\bf 78} (1997) 998,
  \href{http://arXiv.org/abs/gr-qc/9605029}{{\tt arXiv:gr-qc/9605029}}.

\bibitem{Cornish:1996hx}
N.~J. Cornish and J.~J. Levin, ``{The Mixmaster universe: a chaotic Farey
  tale},'' Phys. Rev. D {\bf 55} (1997) 7489,
  \href{http://arXiv.org/abs/gr-qc/9612066}{{\tt arXiv:gr-qc/9612066}}.

\bibitem{Brizuela:2022uun}
D.~Brizuela and S.~F. Uria, ``{Semiclassical study of the Mixmaster model: the
  quantum Kasner map},'' Phys. Rev. D {\bf 106} (2022) 064051,
  \href{http://arXiv.org/abs/2207.00566}{{\tt arXiv:2207.00566}}.

\bibitem{Bojowald:2023fas}
M.~Bojowald, D.~Brizuela, P.~Calizaya~Cabrera, and S.~F. Uria, ``{Chaotic
  behavior of the Bianchi IX model under the influence of quantum effects},''
  Phys. Rev. D {\bf 109} (2024) 044038,
  \href{http://arXiv.org/abs/2307.00063}{{\tt arXiv:2307.00063}}.

\bibitem{Bojowald:2023sjw}
M.~Bojowald, D.~Brizuela, P.~Calizaya~Cabrera, and S.~F. Uria, ``{Reduction of
  primordial chaos by generic quantum effects},'' Phys. Rev. D {\bf 108} (2023)
  L061501, \href{http://arXiv.org/abs/2307.13040}{{\tt arXiv:2307.13040}}.

\bibitem{Matzner:1970}
R.~A. Matzner, L.~C. Shepley, and J.~B. Warren, ``{Dynamics of
  SO(3,R)-homogeneous cosmologies},'' Annals of Physics {\bf 57} (1970) 401.

\bibitem{Ryan:1971}
M.~P. Ryan, ``{Qualitative cosmology: diagrammatic solutions for Bianchi type
  IX universes with expansion, rotation, and shear. I The symmetric case},''
  Annals of Physics {\bf 65} (1971) 506.

\bibitem{Ryan:1971bis}
M.~P. Ryan, ``{Qualitative cosmology: diagrammatic solutions for Bianchi type
  IX universes with expansion, rotation, and shear. II The general case},''
  Annals of Physics {\bf 68} (1971) 541.

\bibitem{Ryan:1972}
M.~P. Ryan, ``{The oscillatory regime near the singularity in Bianchi-Type IX
  universes},'' Annals of Physics {\bf 70} (1972) 301.

\bibitem{Ryanlectures}
M.~P. Ryan, {\em Hamiltonian cosmology (Lecture notes in physics)}.
\newblock Springer-Verlag, Berlin, 1972.

\bibitem{Ryan-Shepley}
M.~P. Ryan and L.~C. Shepley, {\em Homogeneous relativistic cosmologies}.
\newblock Princeton University Press, Princeton, 1975.

\bibitem{Ringstrom:2000mk}
H.~Ringstrom, ``{The Bianchi IX attractor},'' Annales Henri Poincare {\bf 2}
  (2001) 405, \href{http://arXiv.org/abs/gr-qc/0006035}{{\tt
  arXiv:gr-qc/0006035}}.

\bibitem{Heinzle:2009eh}
J.~M. Heinzle and C.~Uggla, ``{A new proof of the Bianchi type IX attractor
  theorem},'' Class. Quant. Grav. {\bf 26} (2009) 075015,
  \href{http://arXiv.org/abs/0901.0806}{{\tt arXiv:0901.0806}}.

\bibitem{Wainwright}
J.~Wainwright and G.~F.~R. Ellis, {\em Dynamical systems in cosmology}.
\newblock Cambridge University Press, Cambridge, 2021.

\bibitem{Jantzen:2001me}
R.~T. Jantzen, ``{Spatially homogeneous dynamics: a unified picture},''
  \href{http://arXiv.org/abs/gr-qc/0102035}{{\tt arXiv:gr-qc/0102035}}.

\bibitem{Heinzle:2009du}
J.~M. Heinzle and C.~Uggla, ``{Mixmaster: fact and belief},'' Class. Quant.
  Grav. {\bf 26} (2009) 075016, \href{http://arXiv.org/abs/0901.0776}{{\tt
  arXiv:0901.0776}}.

\bibitem{Czuchry:2014hxa}
E.~Czuchry, N.~Kwidzinski, and W.~Piechocki, ``{Comparing the dynamics of
  diagonal and general Bianchi IX spacetime},'' Eur. Phys. J. C {\bf 79} (2019)
  173, \href{http://arXiv.org/abs/1409.2206}{{\tt arXiv:1409.2206}}.

\bibitem{Kiefer:2018uyv}
C.~Kiefer, N.~Kwidzinski, and W.~Piechocki, ``{On the dynamics of the general
  Bianchi IX spacetime near the singularity},'' Eur. Phys. J. C {\bf 78} (2018)
  691, \href{http://arXiv.org/abs/1807.06261}{{\tt arXiv:1807.06261}}.

\bibitem{Kwidzinski:2019rwj}
N.~Kwidzinski and W.~Piechocki, ``{Curvature invariants for the Bianchi IX
  spacetime filled with tilted dust},'' Eur. Phys. J. C {\bf 79} (2019) 199,
  \href{http://arXiv.org/abs/1901.01790}{{\tt arXiv:1901.01790}}.

\bibitem{Lin:1990tq}
X.-f. Lin and R.~M. Wald, ``{Proof of the closed universe recollapse conjecture
  for general Bianchi type-IX cosmologies},'' Phys. Rev. D {\bf 41} (1990)
  2444.

\bibitem{Lin:1989tv}
X.-f. Lin and R.~M. Wald, ``{Proof of the closed universe recollapse conjecture
  for diagonal Bianchi type-IX cosmologies},'' Phys. Rev. D {\bf 40} (1989)
  3280.

\bibitem{Kasner:1921zz}
E.~Kasner, ``{Geometrical theorems on Einstein's cosmological equations},'' Am.
  J. Math. {\bf 43} (1921) 217.

\bibitem{Sinai:1970}
Y.~Sinai, ``Dynamical systems with elastic reflections,'' Russian Mathematical
  Surveys {\bf 25} (1970) 137.

\bibitem{Ali:2017qwa}
M.~Ali and V.~Husain, ``{Mixmaster dynamics in the dust time gauge},'' Phys.
  Rev. D {\bf 96} (2017) 044032, \href{http://arXiv.org/abs/1707.07098}{{\tt
  arXiv:1707.07098}}.

\end{thebibliography}\endgroup

\end{document}